%% file: main.tex
\documentclass[final,3p,times,twocolumn]{elsarticle}

\usepackage{amssymb}
\usepackage{amsmath}

\biboptions{longnamesfirst,square,comma,sort&compress}

\usepackage{hyperref}
\usepackage[all]{hypcap} 
\hypersetup{pdfauthor={Name}}
\usepackage[capitalise]{cleveref}

\usepackage{xcolor,soul}

\journal{Astroparticle Physics}

\begin{document}

\begin{frontmatter}

\title{Cosmic-Ray Physics at the South Pole}

\author[1,2]{D.~Soldin\corref{cor1}}
\affiliation[1]{organization={Department of Physics and Astronomy, University of Utah},
            city={Salt Lake City},
            postcode={84112}, 
            state={UT},
            country={USA}}
\affiliation[2]{organization={Karlsruhe Institute of Technology, Institute of Experimental Particle Physics},
            city={Karlsruhe},
            postcode={D-76021}, 
            country={Germany}}

\cortext[cor1]{Corresponding author}

\ead{dennis.soldin@utah.edu}

\author[3]{P.~A.~Evenson}
\affiliation[3]{organization={Bartol Research Institute and Dept. of Physics and Astronomy,
University of Delaware},
            city={Newark},
            postcode={19716}, 
            state={DE},
            country={USA}}

\author[4,5]{H.~Kolanoski}
\affiliation[4]{organization={Institut für Physik, Humboldt-Universität zu Berlin},
            city={Berlin},
            postcode={D-12489}, 
            country={Germany}}
\affiliation[5]{organization={Deutsches Elektronen-Synchrotron DESY},
            city={Zeuthen},
            postcode={D-15738}, 
            country={Germany}}
            
\author[6]{A.~A.~Watson}
\affiliation[6]{organization={School of Physics and Astronomy, University of Leeds},
            city={Leeds},
            country={United Kingdom}}

\begin{abstract}
The geographic South Pole provides unique opportunities to study cosmic particles in the Southern Hemisphere. It represents an optimal location to deploy large-scale neutrino telescopes in the deep Antarctic ice, such as AMANDA or IceCube. In both cases, the presence of an array, constructed to observe extensive air showers, enables hybrid measurements of cosmic rays. While additional neutron monitors can provide information on solar cosmic rays, large detector arrays, like SPASE or IceTop, allow for precise measurements of cosmic rays with energies above several $100\,\rm{TeV}$. In coincidence with the signals recorded in the deep ice, which are mostly due to the high-energy muons produced in air showers, this hybrid detector setup provides important information about the nature of cosmic rays. 

In this review, we will discuss the historical motivation and developments towards measurements of cosmic rays at the geographic South Pole and highlight recent results reported by the IceCube Collaboration. We will emphasize the important contributions by Thomas K. Gaisser and his colleagues that ultimately led to the rich Antarctic research program which today provides crucial insights into cosmic-ray physics.
\end{abstract}


\end{frontmatter}


\section{Introduction}\label{sec:intro}

\input{sec1}

\section{Early years and the role of Bartol}\label{sec:solar_CRs}

\input{sec2}

\section{First air-shower arrays at the South Pole}\label{sec:history}

\input{sec3}

\section{Results from cosmic-ray measurements with the IceCube Neutrino Observatory}\label{sec:HE_CRs}

\input{sec4}

\section{Concluding remarks}\label{sec:conclusions}

\input{sec5}

\section*{Acknowledgements}
The authors would like to gratefully thank the organizers of the Gaisser Memorial Meeting 2023, Newark, DE, USA. H.\,K. and D.\,S. would like to thank the IceCube Collaboration for useful comments and input to this manuscript. A.\,A.\,W. is grateful to Simon Hart, Jim Hinton, Nigel Smith, and Serap Tilav for refreshing his memory on some points of detail.

\bibliographystyle{elsarticle-num} 
\bibliography{main}

\end{document}

%% file: sec1.tex
It was early realized that the geographic South Pole provides opportunities to study cosmic particles in the Southern Hemisphere. In particular, the very low geomagnetic cutoff and high altitude make the South Pole one of the most attractive locations for studies of the production of particles in solar flares.  Work led by Martin Pomerantz of the Bartol Institute began there in 1964 following his earlier pioneering activities at McMurdo. 

Cosmic ray interactions in the atmosphere produce large particle cascades, so-called extensive air showers, and the atmospheric depth at which the maximum of particles is produced scales logarithmically with the initial cosmic ray energy. Thus the high altitude at the South Pole of about $2800\,$m allows for a low energy threshold for shower arrays. In the late 1980s, following the claim for the detection of PeV gamma rays from Cygnus X-3, and X-ray binary~\cite{Hillas:1986}, it was recognised that there was value in establishing an air shower array at the South Pole site. Accordingly, an air-shower array, SPASE, was constructed in 1987 to search for gamma-rays from X-ray binaries and, additionally, for photons from SN1987A which is optimally located for study from this location.  

Subsequently, the SPASE device was combined with the burgeoning activity to construct a neutrino detector with measurements of the mass of cosmic rays being made using a combination of instruments.  In coincidence with neutrino detectors deployed in the deep Antarctic ice, hybrid measurements provide unique information about extensive air showers initiated by cosmic rays with energies above a few $100\,\rm{TeV}$. While a surface array mainly measures the electromagnetic particle content, as well as low-energy muons in an air shower, detectors in the deep ice probe high-energy penetrating muons. This hybrid setup provides important insights into air shower physics and allows to study the nature of cosmic rays in great detail and, along with the gamma ray searches, was a particular interest of Thomas (Tom) Gaisser.

In this report, the motivation and historical developments towards cosmic-ray measurements at the South Pole will be reviewed. In particular, the important contributions to efforts to establish a cosmic-ray science program in Antarctica by the Bartol Research Institute (\cref{sec:solar_CRs}) and the Bartol-Leeds collaboration (\cref{sec:history}), with major contributions by Tom Gaisser, will be discussed. These efforts led to the deployment of the first large-scale neutrino detector in conjunction with an air-shower array at the South Pole, AMANDA and SPASE, and ultimately to the construction of the \emph{IceCube Neutrino Observatory} (IceCube), 
which will be discussed in \cref{sec:HE_CRs}.

%% file: sec2.tex
The first director of the Bartol Research Institute, William F. G. Swann, was widely known for contributions both in theory and experiment. The second director, Martin Pomerantz, carried on the work with ground-based detectors, balloons, and at least one spacecraft. The 1950s introduced the golden age of the neutron monitor, developed by John Simpson of the University of Chicago \cite{PhysRev90934}. Neutron monitors enabled orders of magnitude better resolution from ground-based measurements of low-energy cosmic rays by detecting the hadronic component of the atmospheric cascade. High time resolution data on fluxes of cosmic rays in the few GeV energy regime allowed detailed study of the production of energetic particles in solar flares and the influence of interplanetary magnetic fields on their propagation in the heliosphere – in what was to come to be called the \emph{Solar Wind} \cite{PhysRev104768}. 

The Bartol Research Institute had an important role in this work, largely due to Shakti Duggal, who, until his untimely death was regarded as one of the premier experts in the field. Martin Pomerantz continued the experimental side by establishing neutron monitors in three critical locations. The neutron monitor in Thule, Greenland, the oldest of the monitors operated by the Bartol Research Institute, began collecting data in 1957 as part of the \emph{International Geophysical Year} (IGY). Thule is unique in that its observing \emph{asymptotic direction} is roughly perpendicular to the ecliptic, pointing northward. The asymptotic direction of a particle, entering the atmosphere at a given polar and zenith angle, is the direction from which the particle enters the magnetosphere. A neutron monitor near the Earth’s magnetic equator will detect positively charged particles whose arrival direction is the more eastward, the lower their energy. Thus, the asymptotic direction of a particle depends on its energy.


\subsection{Neutron monitors in Antarctica}
\begin{figure}[!b]
\vspace{0.2em}

\mbox{\hspace{-0.2em}\includegraphics[width=0.49\textwidth]{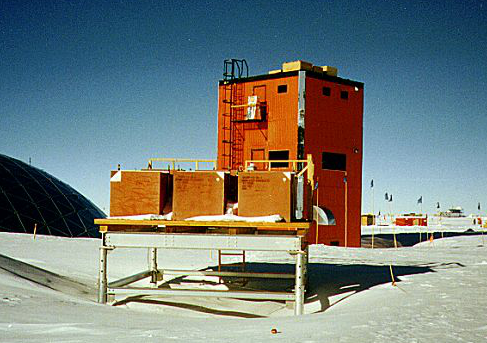}}
\caption{Photo of the three detector tubes in the initial installation of the NM-64 neutron monitors at the South Pole. The neutron monitors are on an elevated platform in insulated, heated boxes. Also shown in the background are the \emph{Skylab} building (housing the monitor electronics, among others) and part of the iconic \emph{Dome} (credit: L.\,Shulman/Bartol).}
\label{fig:PE_1}
\end{figure}

After the IGY, Pomerantz was approached by the National Science Foundation (NSF) regarding the establishment of neutron monitors in Antarctica. The first was established in 1960 at McMurdo Station, which is the geomagnetic complement of Thule in that its asymptotic direction is roughly perpendicular to the ecliptic but pointing southward. In 1964 the original IGY design neutron monitor at McMurdo was replaced by the more modern NM64 design \cite{Hatton1964}, while the IGY monitor was relocated to the South Pole where it has a more equatorial average asymptotic direction but the high altitude and very low geomagnetic cutoff make it still one of the most sensitive ground-based detectors for solar cosmic rays. In 1974 the IGY monitor at the South Pole was replaced by the present NM64 instruments (see {\cref{fig:PE_1}})~\cite{2011ICRCEvenson}.

The Antarctic neutron monitors were in fact the first permanent scientific instrument of their scale deployed on the continent. For many years, the Bartol Research Institute continued a partnership with the NSF, both formally and informally, that resulted in the deployment of many other instruments facilitated by the extensive experience in the logistics and practicality of operation in Antarctica. Through 1985 Bartol had the contract from NSF to provide the winter-over observers who took care of 
the instruments. Even beyond this – indeed to the present day – the Bartol Research Institute has unique experience and capability in Antarctic operations.

In terms of low-energy cosmic rays, the focus of the effort of the Bartol Institute shifted to balloon instruments \cite{Mechbal:2020ela} and the establishment of multiple neutron monitors, with Bartol stations in Nain, Peawanuck, Fort Smith and Inuvik in Canada. Partner stations using Bartol electronics were installed at Syowa and Jang Bogo (Antarctica), Daejeon (Korea), and Doi Inthanon (Thailand). The Bartol Institute and University of Tasmania developed, and for many years operated, the shipborne mobile neutron monitors that are now being used by collaborators at Chiang Mai University in Thailand.

While the neutron monitors continue to operate, the primary technical focus of the Bartol effort at South Pole shifted to higher energy with SPASE, AMANDA, and IceCube with IceTop, as described in the following sections. Although far from the primary objective of the array, the discriminator rates in IceTop (see \cref{sec:detector}) form an excellent complement to the neutron monitor for GeV solar and cosmic particles. In fact, the first letter-class publication from the IceCube Collaboration was a report of the energy spectrum of a solar flare~\cite{IceCube:2008mcx}.

%% file: sec3.tex
The concept of constructing a high-energy air-shower array at the South Pole stemmed from claims made by a group at the University of Kiel in Germany for the detection of gamma-rays of around $1\,\rm{PeV}$ from Cygnus X-3, an X-ray binary, with a local experiment~\cite{Samorski:1983zm}. This result was apparently confirmed by the Haverah Park group \cite{Lloyd-Evans:1983olz}. Signals displaying the $4.8\,\rm{h}$ periodicity of the source, as seen at $1\,\rm{PeV}$, had previously been reported at $1\,\rm{TeV}$ using air-Cherenkov telescopes \cite{Danaher:1981}. Subsequently, other groups reported detections at about $1\,\rm{PeV}$ from several different X-ray binaries, including Hercules X-1. In 1986, Hillas pointed out that if such objects were indeed PeV sources then it would be optimal to search for them at sites in the Southern Hemisphere, as more lay in directions towards the galactic centre \cite{Hillas:1986}. He argued that the South Pole was a particularly advantageous site with the atmospheric depth of about $690\,\rm{g/cm}^{2}$ allowing a low energy threshold for shower arrays, and with potential sources visible for $24$ hours each day at a constant elevation. The downside is that South Pole is a location with a harsh environment.

There were already strong links between people in the Bartol Research Institute, University of Delaware, and in the Department of Physics at the University of Leeds. Tom Gaisser had visited Leeds on several occasions and had collaborated with Michael Hillas, while he and Martin Pomerantz both had a strong personal relationship with Alan Watson, the head of the Haverah Park team.  Accordingly, in mid-1986, a Bartol-Leeds collaboration was formed to operate an air-shower array at the South Pole, the \emph{South Pole Air Shower} \emph{Experiment} (SPASE, later re-named SPASE-1).

Discussion about the project started in late 1986 with deployment beginning in November 1987. The discovery of SN1987A in February 1987 gave added impetus to the efforts, particularly as belief in the credibility of the signals from Cygnus X-3, and from other X-ray binaries, had begun to fade. In addition, there were predictions of high-energy gamma-ray signals from SN1987A to be tested \cite{Gaisser:1987,Berezinsky:1987}.

\begin{figure}[tb]
    \mbox{\hspace{-0.25em}\includegraphics[width=0.485\textwidth]{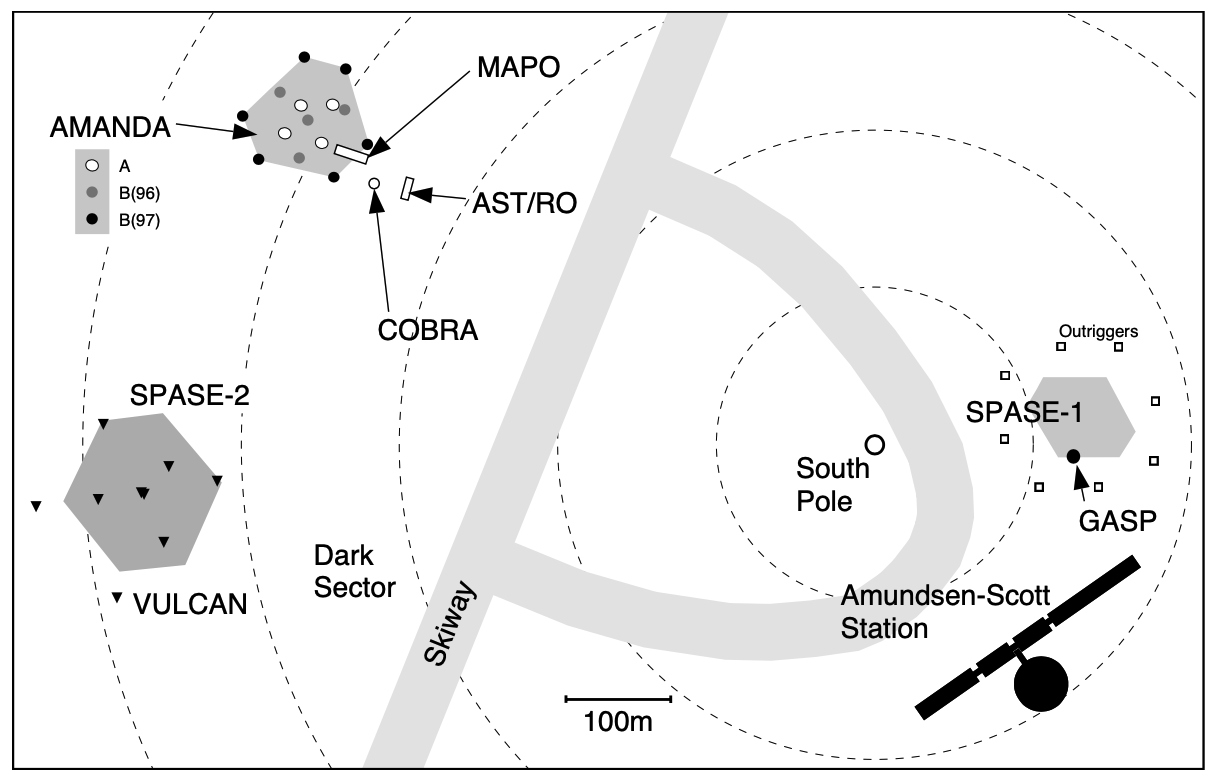}}
    \vspace{-1em}
    
    \caption{The layout of various detector systems at the South Pole in the mid-1990s~\cite{HintonPhD}. Darker gray areas indicate the locations of the SPASE-1 (right), SPASE-II (lower left), and AMANDA (upper left) arrays. Also shown are the Amundsen-Scott Station and the Skiway. GASP was a telescope designed to detect TeV-photons.}
    \label{fig:AW_1}
    \vspace{-.5em}
    
\end{figure}

\subsection{Features of the array and early results}
The Haverah Park group was already operating the \emph{Gamma Ray Experiment} (GREX)\footnote{It was pointed out by Julia Gaisser that the name was an appropriate choice as \textit{grex} is the Latin word for \textit{flock}: flocks of sheep grazed on the Haverah Park array.} array of $32$ scintillators with an area of $1\,\rm{m}^2$ each \cite{Brooke:1985} to explore the Cygnus X-3 claims further, and the design for an array at the Pole was based upon it. The positions of the SPASE array, and of other instruments deployed subsequently, are shown in \cref{fig:AW_1}. Fourteen detectors of SPASE-1 were deployed on a $30\,\rm{m}$ triangular grid, with additional detectors placed adjacent to two of them to allow an empirical determination of the accuracies of timing and signal measurements. Details of the array and of the performance are given in Ref.~\cite{Smith:1989}. For several years, the work was supported by the NSF Division of Polar Programs and by the University of Leeds, with funding coming later from the UK Particle Physics and Astronomy Research Council.

\begin{figure}[tb]
    \centering
    \includegraphics[width=0.478\textwidth]{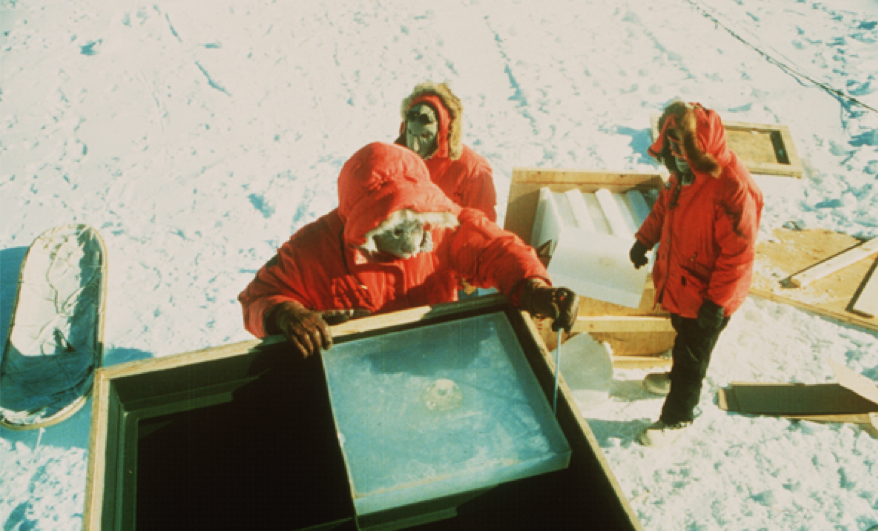}
    \vspace{-1.2em}
    
    \caption{Loading a block of scintillator for the SPASE air-shower array at the South Pole (credit: University of Leeds Collection).}
    \label{fig:AW_21}
    \vspace{-1em}
\end{figure}

Plastic scintillators, 
loaned by John Linsley, were mounted in $1\,\rm{m}^3$ cubical boxes (see \cref{fig:AW_21}) and viewed from below with a $3\,\rm{inch}$ EMI 9821B photomultiplier. As these photomultipliers had not been operated at the low temperatures experienced at the Pole, a study of their performance was undertaken in Leeds with tests made down to $-80^\circ$C \cite{McMillan:1989}. The wooden boxes used to house the scintillators had been manufactured in McMurdo out of rather poor quality wood and were far from light-tight, but one of the on-site carpenters, Ray Burdie, successfully dealt with this problem.

Monitoring the stability of the timing of the signals from the detectors was essential. Light pulses with very short rise time generated by a laser controlled by a Pockels cell, which were also used at Haverah Park, were linked to each detector with optical fibers and incorporated in the SPASE design. Cable lengths were measured to better than $0.1\,\rm{ns}$ using a time-domain reflectometer and checked using Fidecaro’s method \cite{Fidecaro:1960}. The cables to each detector were buried under approximately $0.3\,\rm{m}$ of snow to minimize the effects of temperature variations. After one year of successful operation, $6\,\rm{mm}$ lead sheets were placed above each detector to improve the angular resolution, as had been demonstrated at Haverah Park \cite{Bloomer:1988}. The lead sheet had three effects: \textit{(i)} low-energy electrons and photons are absorbed and no longer contribute to the signal, \textit{(ii)} high-energy electrons produce an enhanced signal size through multiplication, \textit{(iii)} high-energy photons materialise, producing additional signal contributions similar in size to those produced by electrons. The number of particles gained from the latter two processes exceeds that lost due to absorption and the so-called \emph{Rossi transition effect}~\cite{Rossi1932} is observed. The enhanced signal reduces the timing fluctuations, improving the angular resolution.  

It was a surprise to discover that the direction of the Greenwich Meridian was not known, as this is an essential to translate the local direction of the showers to celestial coordinates. A sundial method was developed to solve this problem, with the measured orientations later checked using the transit of the Sun along lines of detectors. After applying this procedure, the orientation of the array and the direction of the Meridian were known to $0.2^\circ$. The distance between detectors was measured with a theodolite and a Hewlett-Packard distance meter.

The array was brought into operation within about six weeks of the team\footnote{The installation team comprised Nigel Smith (who was the first to winter-over in support of the SPASE project), Jay Perrett, Paul Ogden and Alan Watson, all from Leeds University. Martin Pomerantz provided a key link to technical staff at the South Pole station.} arriving at the Pole and air showers were recorded at a rate of around $0.6 \times 10^6$ per week. The equipment worked flawlessly throughout the winter with an on-time $>95\%$, the only surprise being the discovery that the electrical length of the signal cables, which were measured regularly, decreased as the temperature fell. This is due to the fall in the dielectric constant in the cable core as the temperature decreases. Over $2 \times 10^6$ events had been analysed before the February closure of the Pole station for the winter. 

During the following summer season (1988/1989), Trevor Weekes installed instrumentation to observe Cherenkov light produced by showers in the atmosphere, with two Fresnel lens/photomultiplier combinations sited $11\,\rm{m}$ from the centre of the SPASE array. The orientation of the Cherenkov telescopes was determined during the winter from a lunar transit where the moon was observed. With this device, it was demonstrated that the angular accuracy of the reconstructed shower direction was around $0.8^\circ$ at $200\,\rm{TeV}$~\cite{Walker:1991}.

At the time of operation of SPASE-1, very little information could be transmitted during the Antarctic winter so magnetic tapes with data from January and February 1988 were brought out when the station closed. Some \mbox{$2 \times 10^6$} events were used to set limits on signals above $100\,\rm{TeV}$ during an interesting epoch when SN1987A was active in X-rays, and when an observation of TeV photons had been reported~\cite{PhysRevLett.61.2292}. No signals were detected with a limit on the output of \mbox{$3 \times 10^{39}\,\rm{erg\,s}^{-1}$} at a median energy of $100\,\rm{TeV}$ \cite{Gaisser:1988sd} which was comparable to the best existing limit, despite the limited information transmitted from South Pole during this time. Using four years of data, a systematic search was made for signals from 9 prospective sources including $6$ X-ray binaries, SN19987A and Tuc47. Only upper limits were obtained \cite{PhysRevD.48.4504}. While the SPASE project was technically very successful \cite{PhysRevD.48.4495}, the results obtained from it alone were scientifically disappointing.

\subsection{Coincident operation with AMANDA detectors}\label{subsec:AMANDA-SPASE}
As a first step towards what became IceCube, a three-photomultiplier string was deployed at a depth of $210\,\rm{m}$ in ice in Greenland, an effort led by Bob Morse (University of Wisconsin), with muons detected at the expected rate \cite{Lowder:1991}. Subsequently, a similar telescope with $4$ photomultipliers was installed at a depth of $800\,\rm{m}$ near the center of the SPASE array and coincidences of muons with air showers were observed in $117$ events \cite{Tilav:1993}.

\begin{figure}[tb]
    \mbox{\hspace{-1.2em}\includegraphics[width=0.52\textwidth]{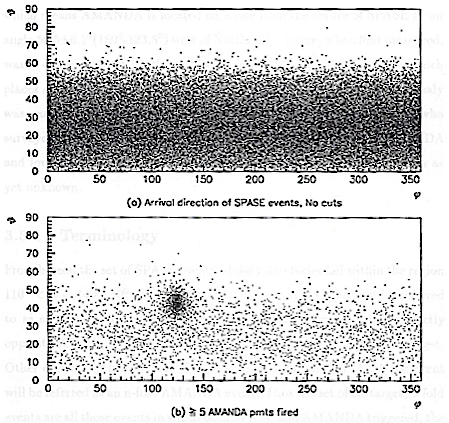}}
    \vspace{-1.5em}
    \caption{The distribution in zenith angle (y-axis) and azimuth angle (x-axis) of events recorded by the SPASE array alone (upper plot) and events recorded when the SPASE array and 5 photomultipliers in the four AMANDA strings were triggered. Figure taken from Ref.~\cite{HartPhD}.}
    \label{fig:AW_2}
    \vspace{-.5em}
\end{figure}

Tom Gaisser had long been interested in the science that could be gleaned from a surface array operated in coincidence with a deep underground detector. His idea was to extract information on mass composition by detecting high-energy muons near to the cores of air showers in coincidence with showers recorded at the surface. Accordingly, soon after the first \emph{Antarctic Muon And Neutrino Detector Array} (AMANDA, later re-named AMANDA-A) was deployed, an effort again led by Bob Morse, coincidences were recorded between SPASE and AMANDA. AMANDA-A comprised four strings, each with $20$ photomultipliers, deployed at depths between $800$ and $1000\,\rm{m}$: photomultipliers in the top layer were pointed upwards. The first results, obtained in February 1994, taken from the PhD thesis of Simon Hart~\cite{HartPhD} who pushed through the initial effort, are shown in \cref{fig:AW_2}. Further details are given in Refs.~\cite{Miller:1995,SPASE:1995zuu}.

These measurements had important consequences. The rate of coincidences between SPASE and AMANDA was much higher than expected and it was quickly realized that Cherenkov light was being scattered by air bubbles in the ice and that the attenuation length of the light was much longer than had been anticipated \cite{AMANDA:1994ttq}. On the one hand, this meant that the IceCube detector (see \cref{sec:HE_CRs}) would have to be constructed at a deeper level, where the bubble density was known to be lower, but that the longer attenuation length ($\sim 400\,\mathrm{m}$) would allow the photomultipliers to be separated by a greater distance.


The pioneering coincidence study had other consequences. It was decided to build a second version of SPASE (SPASE-II) closer to AMANDA, which was enhanced as AMANDA-B10 \cite{AMANDA:2002pgr}. Tom Gaisser was heavily involved in the planning of SPASE-II\footnote{Alan Watson recalls asking Tom Gaisser to estimate and cost the cables for the project. He was delighted to do this and said that \emph{``this was the first time that he’d had to do a calculation that actually meant something!''}. Perhaps this led him to experimental work in which he was later so active during visits to the South Pole to work on IceCube.} where the detectors were again made of plastic scintillators but with a different design \cite{DICKINSON200095}. The SPASE-II array consisted of $30$ detector stations, $30\,\rm{m}$ apart on a triangular grid, with $5$ detector modules each. Each detector module housed a $1\,\rm{cm}$-thick hexagonal piece of plastic scintillator with triangular light guides in the center which redirected the light to a $2\,\rm{inch}$ photomultiplier. An array of air-Cherenkov detectors (VULCAN) was also constructed and used successfully to measure the depth of the maximum of air showers in the energy range of about $1-10\,\rm{PeV}$ \cite{DICKINSON2000114,Dickinson:1999}.

The combination of the two SPASE detectors and AMANDA-B10 proved to be a powerful one. A detailed survey of the positions of the photomultiplier tubes under the ice was made, the work being led by Tom Gaisser, Tom Miller, and Serap Tilav \cite{SPASE:2004bne}. One result of studies of the muon tracks in the deep ice was the discovery that the location of one AMANDA-A string, as provided by the deployers, was out by $10\,\rm{m}$. 

The main science result from the coincidence work was a measurement of the mean mass of cosmic rays in the range $500\,\rm{TeV}$ to $5\,\rm{PeV}$. Over this range, the mean value of $\langle \ln\,A\rangle$, where $A$ is the atomic mass, was found to increase by $0.8$ \cite{AMANDA:2004bgc}. Thus, one of Tom Gaisser’s many science goals was achieved and the successful coincident operation of AMANDA and SPASE detectors laid the foundations towards the subsequent construction of the IceTop surface array as part of the IceCube Neutrino Observatory.

%% file: sec4.tex
The construction of the IceCube Neutrino Observatory at the South Pole began in 2004, after the first NSF proposal had been submitted in 1999. Dedicated teams steadily deployed parts of the detector each year from November to February until the end of 2010. On December 18, 2010, just after 6 pm New Zealand time, the final part of the detector was commissioned and IceCube was completed, only a decade after the collaboration submitted the first proposal. Ever since, IceCube is reporting a large variety of exciting science results, which are discussed below\footnote{This section is mostly based on conference proceedings prepared by the current authors, in particular Refs.~\cite{Kolanoski:2022klq,Soldin:2018vak,Soldin:2019mpv}.}.

Although the main focus of the detector was the discovery of astrophysical neutrinos, Tom Gaisser played a major role ensuring that this new detector becomes a multi-purpose experiment with unique opportunities to study cosmic rays. In particular, coincident measurements of air showers at the surface and the accompanying high-energy muons in the South Pole ice were of large interest. The concept had already been successfully shown by the coincident operation of the AMANDA and SPASE detectors.

\subsection{The detector}\label{sec:detector}
\begin{figure}[tb]
    \vspace{-1em}
    \centering
    \includegraphics[width=0.42\textwidth]{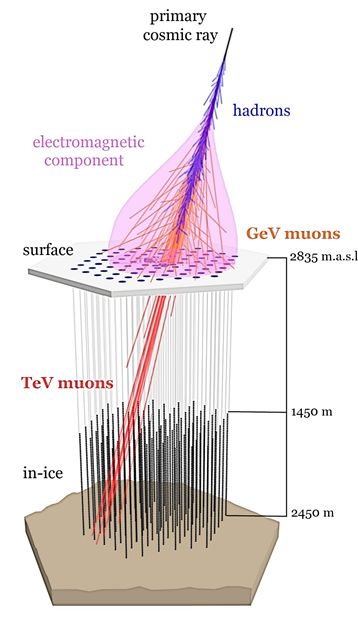}
    \vspace{-0.5em}
    \caption{The IceCube Neutrino Observatory~\cite{Aartsen:2016nxy}. Shown are the in-ice detector and the surface array IceTop (see text for details).}
    \label{fig:IC-IT}
\end{figure}

The IceCube Neutrino Observatory is a cubic-kilometer detector situated deep in the ice at the geographic South Pole, at a depth of about 2000\,m~\cite{Aartsen:2016nxy}, as shown in \cref{fig:IC-IT}. IceTop, the surface component of IceCube, is an air-shower array covering the energy range from about $10^{14}$\,eV to $10^{18}$\,eV~\cite{IceTop}, that is the energy range between direct measurements with balloons and satellites and the highest energies covered by experiments like the Telescope Array (TA) and Pierre Auger Observatory (Auger). Besides serving for the physics of charged cosmic rays, the surface array is also employed as veto against cosmic-ray induced background in the search for astrophysical neutrinos with IceCube~\cite{Tosi:2019nau}. 

As previously described, Tom Gaisser had a very strong interest in measurements of cosmic rays with a surface array operated in coincidence with a deep underground detector. This is reflected in the design of IceCube in that cosmic-ray physics is performed with the air-shower array IceTop as well as with the in-ice detector IceCube, both independently or together in coincidence. While IceTop exploits the air showers, the in-ice detector provides the detection of high-energy muons or muon bundles. Coincidence measurements are the particular strength of IceCube, supplying a powerful handle for the determination of the mass composition and unique tests for air-shower physics, for example. 

\subsubsection{IceCube}\label{sec:icecube}
IceCube's main component is an array of 86 strings equipped with 5160 \emph{Digital Optical Modules} (DOMs) in a volume of 1\,km$^3$ at a depth between 1450\,m and 2450\,m \cite{Aartsen:2016nxy}. Each DOM contains a photomultiplier tube to record the Cherenkov light of charged particles that penetrate the ice. In addition, a DOM houses electronics supplying signal digitization, readout, triggering, calibration, data transfer, and various control functions. In the central lower part of the detector, a section called DeepCore is more densely instrumented. 

The main purpose of IceCube is the detection of high-energy neutrinos from astrophysical sources via the Cherenkov light of charged particles generated in neutrino interactions in the ice or the rock below the deep ice. In the context of cosmic-ray physics it allows for the detection of high-energy muons and neutrinos generated in extensive air showers initiated by cosmic rays in the atmosphere.

\subsubsection{IceTop}\label{sec:icetop}

\begin{figure}[tb]
    \vspace{0.1em}
    \mbox{\hspace{-.5em}\includegraphics[width=0.49\textwidth]{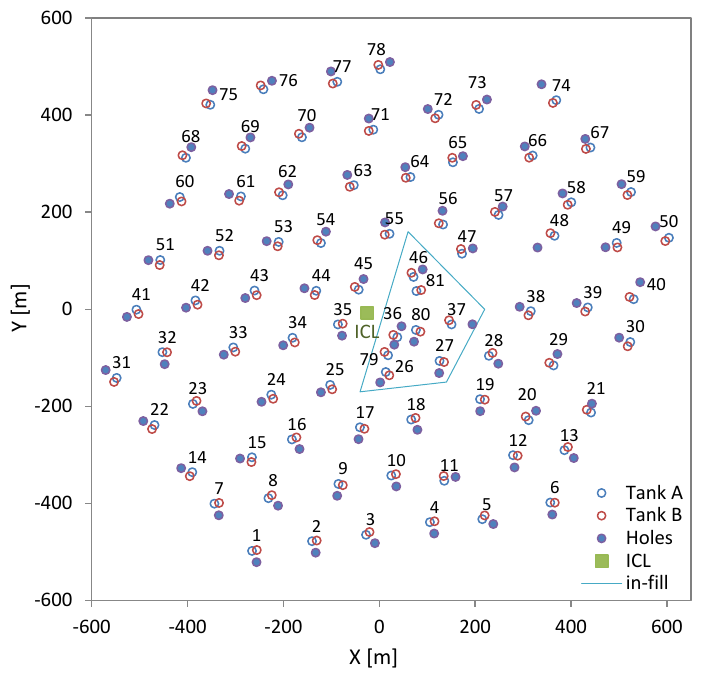}}
    \vspace{-1.5em}
    
    \caption{Surface array IceTop~\cite{IceTop}, including a denser instrumented in-fill region. Shown are the two tanks (A/B) comprising a station, the location of the in-ice strings (Holes), and the \emph{IceCube Lab} (ICL).}
    \label{fig:IT-array}
    \vspace{-0.8em}
\end{figure}
\begin{figure}[tb]
    \vspace{.7em}
    \centering
    \includegraphics[width=0.47\textwidth]{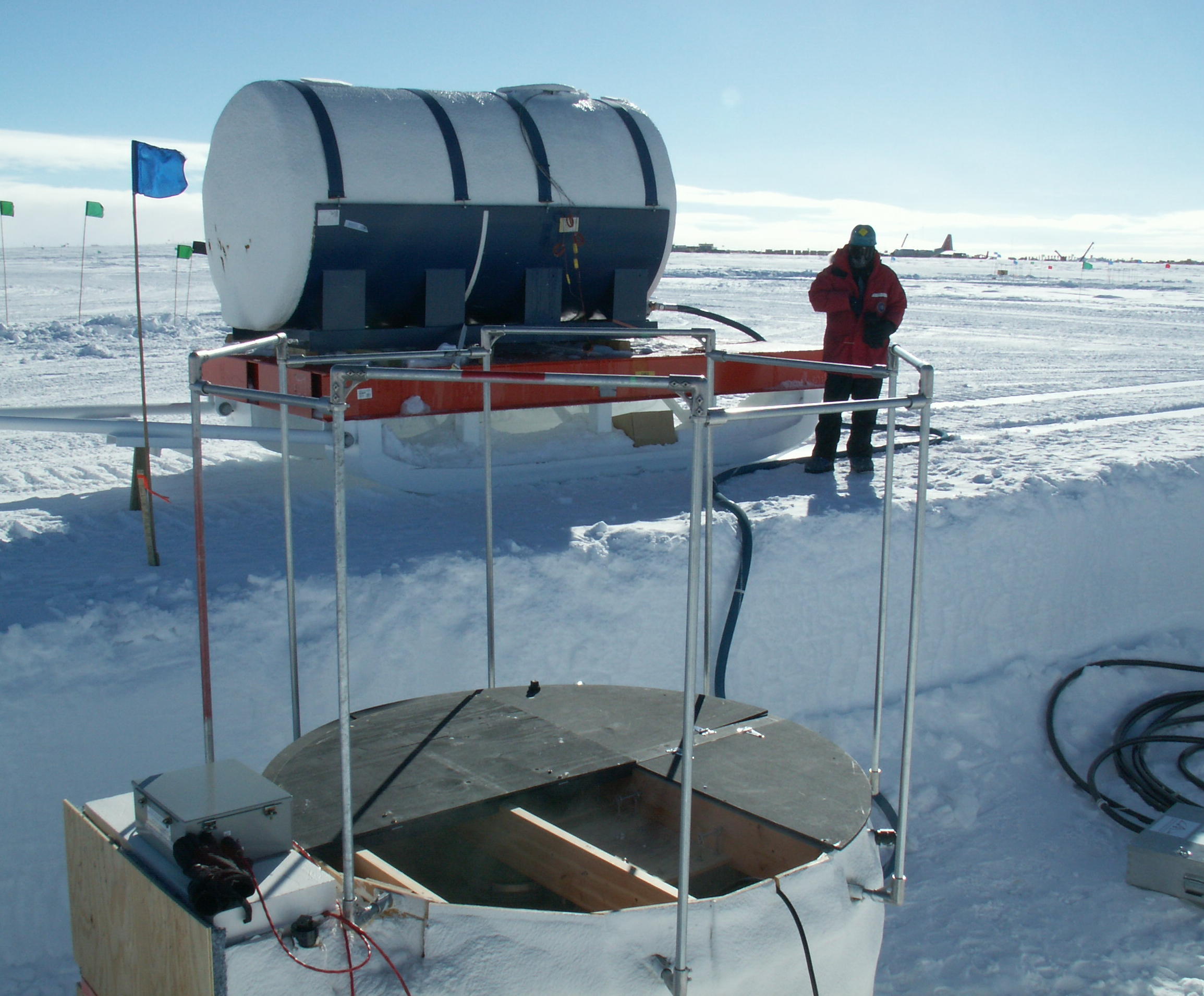}
    
    \vspace{.5em}
    
    \includegraphics[width=0.47\textwidth]{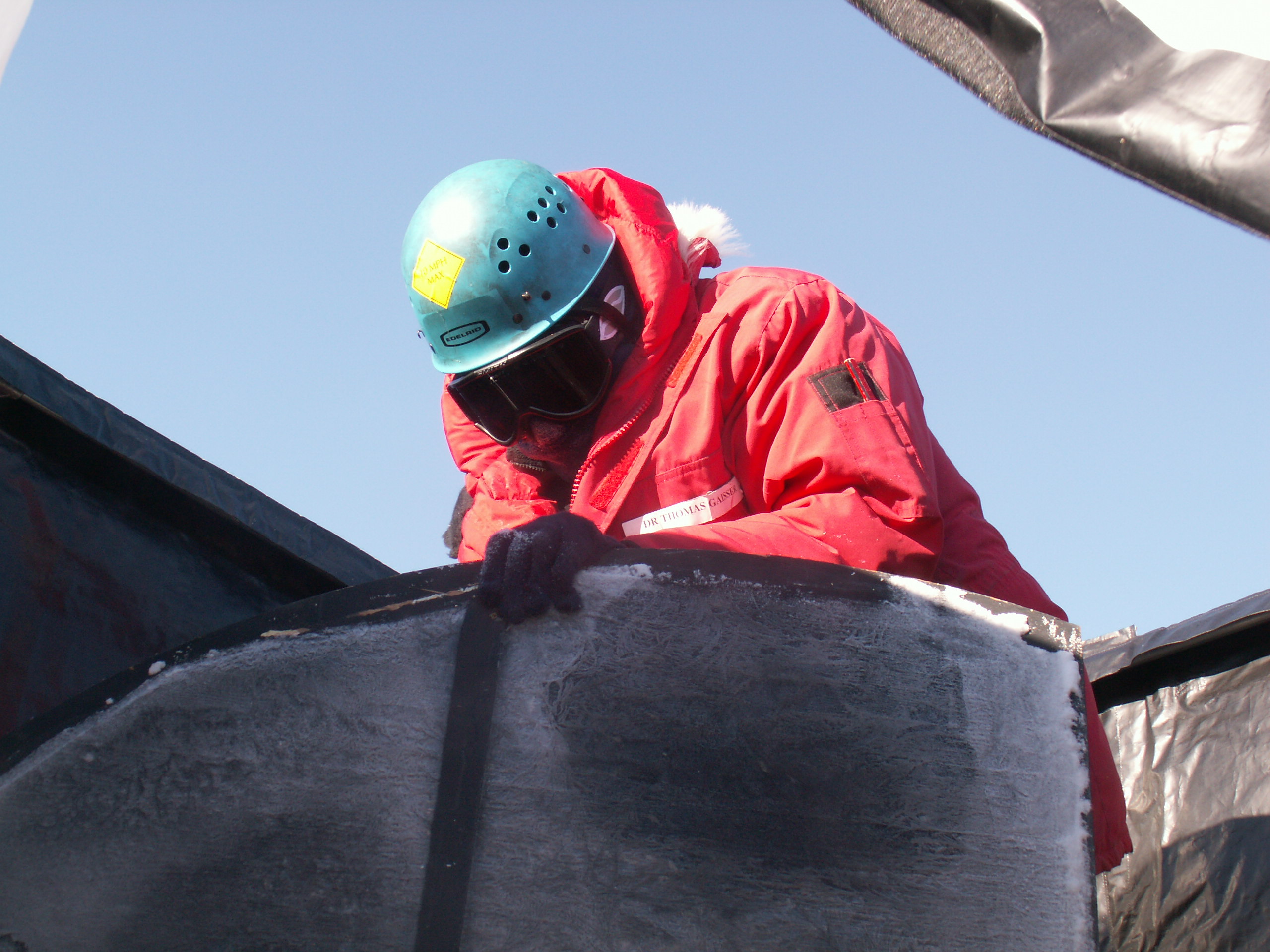}
    \caption{Tom Gaisser supervising and monitoring the filling of an IceTop tank with water for the deployment of the surface array during the austral summer of 2008/2009 (credit: H.\,Kolanoski/NSF).}
    \label{fig:Tom_at_Pole}
\end{figure}

\begin{figure*}[tb]
    \vspace{-3.em}
    \centering
    \includegraphics[width=0.82\textwidth]{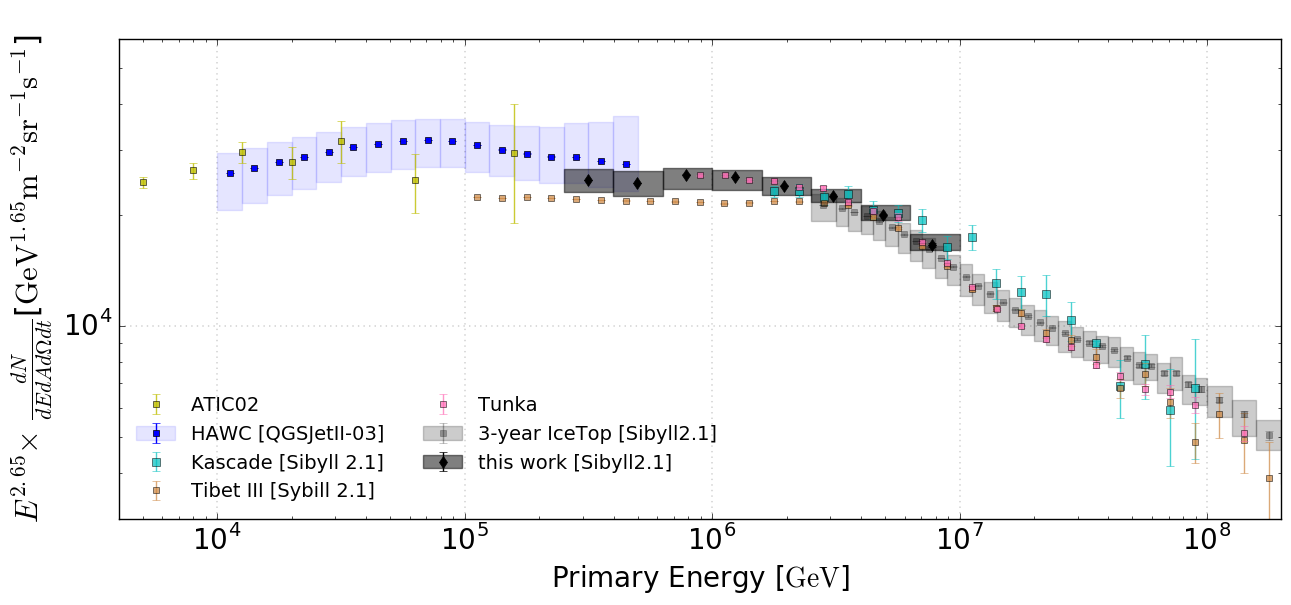}
    \caption{Cosmic-ray flux (differential in $\ln E$) using IceTop data scaled by $E^{1.65}$ and compared with fluxes from other experiments (references in Ref.~\cite{IceCube:2020yct}). Note that different hadronic interaction models are used by the experiments as indicated by the labels in the square-brackets.}
    \label{fig:all-particle-spectra-IT}
\end{figure*}

Tom Gaisser was not only an excellent theorist, but he was also well acquainted with detector hardware -- IceTop is his child. Together with his group at the Bartol Research Institute, he developed the detector concept for IceTop and initiated the manufacturing of the ice-Cherenkov tanks. The IceTop air-shower array~\cite{IceTop} is located above IceCube at a height of 2832\,m above sea level, corresponding to an atmospheric depth of about 690\,g/cm$^2$. It consists of 162 ice-Cherenkov tanks, placed at 81 stations and distributed over an area of 1\,km$^2$ on a grid with mean spacing of 125\,m, as shown in \cref{fig:IT-array}. In the center of the array, three stations have been installed at intermediate positions which was a special request by Tom Gaisser who wanted to cover a maximal cosmic-ray energy range. Together with the neighboring stations, they form an in-fill array for denser shower sampling allowing for lower energy thresholds. Each station comprises two cylindrical tanks, $10\,\rm{m}$ apart from each other,  with a diameter of $1.86\,\rm{m}$ and filled with $90\,\rm{cm}$ of ice. \Cref{fig:Tom_at_Pole} shows the filling of an IceTop tank with water supervised by Tom Gaisser. The tanks are embedded into the snow so that ideally their top surface is level with the surrounding snow to minimize temperature variations and snow accumulation caused by wind drift. However, snow accumulation (mainly due to irregular snow surfaces) cannot be avoided so that the snow height has to be monitored and taken into account in simulation and reconstruction. 

Each tank is equipped with two DOMs, where the electronics and readout scheme are the same as for the in-ice detector \cite{Aartsen:2016nxy}, as previously described.
To initiate the readout of DOMs, a local coincidence between the tanks of a station is required. Additionally, IceTop is always read out in case of a trigger issued by another detector component (and vice versa). For each single tank above threshold, even without a local coincidence, condensed data containing integrated charge and time stamp are transmitted. These so-called SLC (\emph{Soft Local Coincidence}) hits are useful for detecting single muons in shower regions where the electromagnetic component has mostly or fully been absorbed, for example for low energy or inclined showers, in the outer region of showers, or a combination of these. For IceTop, the measured charges are expressed in units of \emph{Vertical Equivalent Muons} (VEM) determined by calibrating each DOM with muons.

\subsection{All-particle spectrum (IceTop only)} \label{sec:All-Particle-Spectrum}

The determination of the spectrum and mass composition of the charged cosmic rays is a key topic of IceCube's cosmic-ray program. The IceCube Collaboration published several analyses of the spectrum which are different in methods and/or in the covered energy range. Analyses done with IceTop only, without the information of the high-energy muons detected in the deep detector, allow for a wider zenith-angle range since a coincidence is only possible for zenith angles smaller than about $30^{\circ}$. The signals recorded by the surface tanks yield the energy deposited by the shower particles together with the arrival times. This information is used to reconstruct the shower energy and direction by fitting the lateral shower distribution around the shower axis. The shower axis is mainly determined by the arrival times of the signals. 
The  lateral distribution of the tank signals, $S(R)$, at a distance R from the shower axis is fitted by the \emph{Lateral Distribution Function} (LDF),
\begin{equation} \label{eq:LDF}
    S(R) = f_{\rm snow}\, S_{\!125} \, \left(\frac{R}{125\,\rm{m}}\right)^{-\beta-0.303\log_{10}\left(\frac{R}{125\,\rm{m}}\right)}	\, ,
\end{equation}
which is equivalent to describing the logarithm of the tank signals as a second order polynomial in the logarithm of $R$.  The shower size is characterized by the signal $S_{\!125}$ at a reference radius of 125\,m. $\beta$ is the slope of the logarithm of the LDF and the function $f_{\rm snow}$ accounts for the signal attenuation due to snow coverage.

The shower size parameter $S_{\!125}$ is used as energy proxy. Although it is chosen to minimize dependencies on other parameters, like the mass of the primary, the relation between $S_{\!125}$ and the energy of the primary cosmic ray has a slight mass dependence. Since in the IceTop-only approach, one does not directly determine the mass, one has to use a model for the mass composition. We mostly refer to Tom Gaisser's H4a model~\cite{Gaisser:2011klf}. Consistency of the model with the data can be checked by requiring that the same spectrum is obtained in all zenith angle directions since the shower development and absorption depend on the slant depth in the atmosphere and differ for different masses of the primaries.

\begin{figure*}[t]
    \vspace{-3em}
    \mbox{\hspace{0.1 cm}\includegraphics[width=0.98\textwidth]{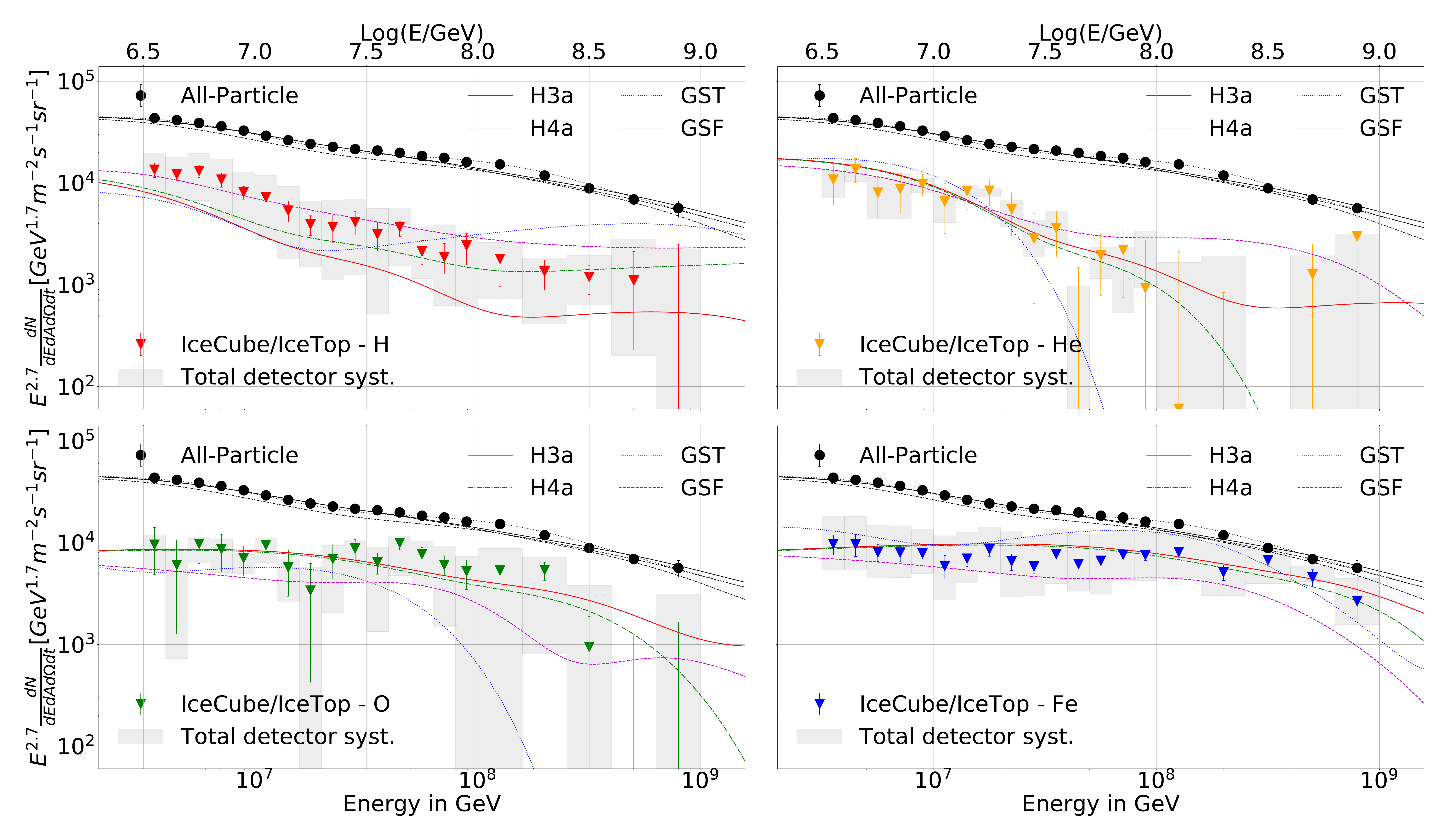}}
    \vspace{-1em}
    \caption{Elementary mass spectra obtained from three years of data from IceCube and IceTop for four elemental mass groups~\cite{IceCube:2019hmk}. Also shown are the systematic uncertainties described in the text and for comparison the predictions from the flux models H3a~\cite{Gaisser:2011klf}, GST~\cite{Gaisser:2013bla}, and GSF~\cite{Dembinski:2017zsh}.} 
    \label{fig:CRMass2}
\end{figure*}

IceCube's latest results on the all-particle spectrum using IceTop data only are shown in \cref{fig:all-particle-spectra-IT}. The plot combines two IceTop analyses, one for energies from 2.5\,PeV to 1.26\,EeV \cite{IceCube:2019hmk} (only shown up to 200\,PeV in the figure) and the other for energies from 250\,TeV to 10\,PeV \cite{IceCube:2020yct}. The latter analysis exploits, for the first time, IceTop infill stations extending IceTop measurements to lower energies and was developed in Tom Gaisser's group at Bartol. In this way the all-particle spectrum of IceTop connects at low energies to direct measurements and data from HAWC, a surface detector at very high altitude \cite{HAWC-Tepe:2012kz}, which was one of Tom Gaisser's major science goals. The uncertainties are mostly dominated by systematics. 

The two IceTop analyses, which use quite different trigger conditions and reconstruction methods, agree within their systematic uncertainties in the overlap region. At low energies they connect well with the HAWC data and the direct measurements by the balloon experiment ATIC02 (see Ref.~\cite{IceCube:2020yct} for references). All spectra that cover the PeV region clearly confirm the knee feature around 4\,PeV. In an earlier publication using less data \cite{IceCube:2013ftu}, IceCube found that between the knee and 1\,EeV, the spectrum exhibits a clear deviation from a single power law. The spectral index changes from \mbox{$\gamma\approx -2.63$} below the knee to about \mbox{$\gamma\approx -3.1$} above the knee, hardens around 18\,PeV towards \mbox{$\gamma\approx -2.9$} and steepens again around 130\,PeV reaching \mbox{$\gamma\approx -3.4$} \cite{IceCube:2013ftu}.

\subsection{Cosmic-ray spectrum and mass composition} \label{sec:theRoleofMuons}

\subsubsection{Mass-sensitive observables in air showers}\label{subsec:mass-sensitivity} 
In the high-energy regime where cosmic rays cannot directly be measured, the mass composition of the primaries can only be inferred from the shower development in the atmosphere. All methods of mass determination are based on the model that a nucleus of mass number $A$ and energy $E$ shares its energy about equally between the $A$ nucleons, hence the energy of nucleon \mbox{$i\ (i=1,\cdots, A$)} is
\begin{equation} \label{eq:nucleonenergy}
    E_i = \frac{E}{A}\, .
\end{equation}
At high energies, the nucleons can be assumed to interact independently, so that one has $A$ independent shower developments. This yields various shower parameters to become dependent on the mass of the primary. For example, a shower composed of many sub-showers, which each have lower energies than the primary, reaches the shower maximum earlier the more sub-showers contribute, thus the shower maximum is mass-dependent. Also, since the ratio of decay and interaction probability of mesons is energy-dependent, the muon multiplicity in an event becomes mass-dependent. While experiments at the highest energies, above about $10^{17}\,\mathrm{eV}$, typically use a measurement of the shower maximum for the determination of the mass composition, IceCube uses the muon multiplicity to determine the cosmic ray mass composition. Mostly muons are measured by the surface detectors. However, IceCube has the additional opportunity to measure high-energy muons in the TeV range (stemming from the first interactions in the atmosphere) and correlate their number per event with the (mainly electromagnetic) shower energy  deposited in the surface detectors.

The energy dependence of the muon multiplicity can be approximated by a power law with an index \mbox{$\beta\approx 0.9$}~\cite{Kampert:2012mx}
such that the muon number per event becomes
\begin{equation} \label{eq:muonrate}
    N_{\mu} \propto A \left(\frac{E}{A}\right)^{\beta} = A^{1-\beta} E^{\beta}\, .
\end{equation}
IceCube can measure multiplicities of TeV-muons in the deep ice as well as of GeV-muons with the surface detectors and can compare these muon counts to the electromagnetic shower component of an event as determined by the surface detector. In the case of GeV-muons this comparison has only been done statistically averaging over many events. From the correlations, the mass composition can be deduced as will be shown in the following.

\subsubsection{Cosmic-ray mass composition} \label{subsec:TeV_muons}

As already emphasized, a strength of IceCube is the possibility to measure high-energy muons in the deep ice in coincidence with the shower reconstructed in IceTop, as indicated in \cref{fig:IC-IT}. This hybrid detector design was strongly influenced by Tom Gaisser's early work related to SPASE and AMANDA coincident measurements (see also \cref{subsec:AMANDA-SPASE}). The muon bundle shown in the figure is narrower than the distance between the strings carrying the optical modules and thus individual muons cannot be resolved. Therefore, instead of the muon count, one exploits the energy deposited by the bundle and the spatial fluctuations of the deposition to get a handle on the muon number. High-energy muons (above the critical energy around 1\,TeV) show strong deposition fluctuations  along their trajectory due to bremsstrahlung. While the ionization energy loss occurs relatively smoothly and nearly energy independently, the energy deposition due to bremsstrahlung occurs more stochastically and is linearly increasing with energy. The surface detectors provide a calorimetric measurement of mainly the electromagnetic component of a shower, depending on the energy, mass and zenith angle of the primary particle. 

\begin{figure}[tb]
   \vspace{-1.4em}
    \mbox{\hspace{-1em}%
    \includegraphics[width=0.48\textwidth]{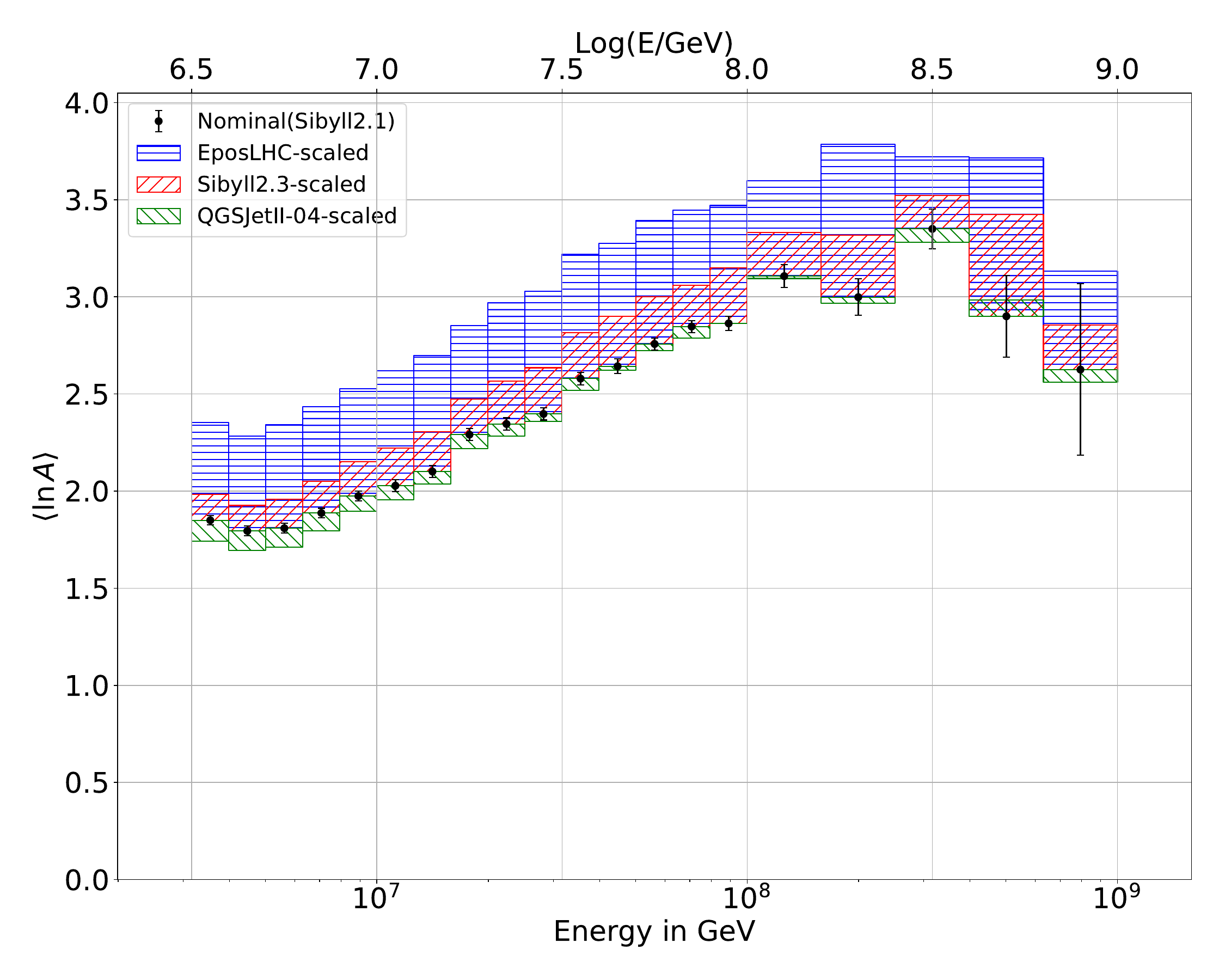}}
    \vspace{-1.em}
    
    \caption{Determination of the average logarithmic mass of the primaries using, as baseline for the training of the network~\cite{IceCube:2019hmk}, the hadronic interaction model Sibyll\,2.1. The bands indicate the shift of the data points if another model is employed.}
    \label{fig:lnA_had_sys}
    \vspace{-1.2em}
\end{figure}

In the latest analysis of the spectrum and mass composition of cosmic rays \cite{IceCube:2019hmk} IceCube uses a neural network in order to exploit as many mass dependent correlations as possible. The neural network has five inputs, two from the surface measurements (energy proxy $S_{\!125}$ and zenith angle) and three from the in-ice measurements (an energy proxy and two \emph{stochasticity} variables). The network is trained to deliver energy and mass of the primary. The training sample uses simulations of hydrogen, helium, oxygen, and iron.

The analysis yields the energy spectra for the four element groups, as shown in \cref{fig:CRMass2}. The various sources of systematic uncertainties are discussed in Ref.~\cite{IceCube:2019hmk}. The sum of these spectra is the all-particle spectrum (\cref{sec:All-Particle-Spectrum}) which agrees well with the IceTop-alone spectrum. Also show in \cref{fig:CRMass2} are the flux model predictions from Refs.~\cite{Gaisser:2011klf,Gaisser:2013bla,Dembinski:2017zsh}. From these individual elemental energy spectra, one can also determine the average mass, conventionally one reports the average logarithm of the mass, $\langle\ln A\rangle$, as a function of energy. \Cref{fig:lnA_had_sys} shows the corresponding results for $\langle\ln A\rangle$ including the dominating uncertainty due to the hadronic interaction model used for the network training. This uncertainty affects both the shower reconstruction as well as the predicted muon multiplicities per event. The effect of hadronic interaction models in the interpretation of data is further discussed in \cref{subsec:Test_hadronic_models} below. 

\subsection{Tests of hadronic interaction models} \label{subsec:Test_hadronic_models}
For the interpretation of air-shower measurements, a correct modeling of the hadronic interactions of shower particles in the atmosphere is crucial. Various  models 
are available which, however, in general yield different results for the same measurement. Therefore, it is important to test the validity of the models and obtain insights for improvements. There are new model versions available which have been updated using LHC data. Thus, one distinguishes between \emph{pre-LHC} models, e.g., Sibyll 2.1~\cite{Ahn:2009wx}, which was developed in the 1990s by Tom Gaisser and his colleagues at Bartol, and more recent \emph{post-LHC} models, such as Sibyll 2.3~\cite{Fedynitch:2018cbl,Riehn:2019jet}, EPOS-LHC~\cite{Pierog:2013ria,Pierog:2015ifw}, and QGSJet-II.04~\cite{Ostapchenko:2005nj,Ostapchenko:2013pia}, or DPMJet-III~\cite{Ranft:2002rj,Roesler:2000he}. Due to its hybrid detector setup, IceCube with IceTop is able to measure the muon content separately in the GeV and TeV energy regimes which provides unique tests of these models.

\subsubsection{GeV-muons} \label{subsec:GeV_muons}

\begin{figure}[tb]
    \mbox{\hspace{-1.em}%
    \includegraphics[width=0.478\textwidth]{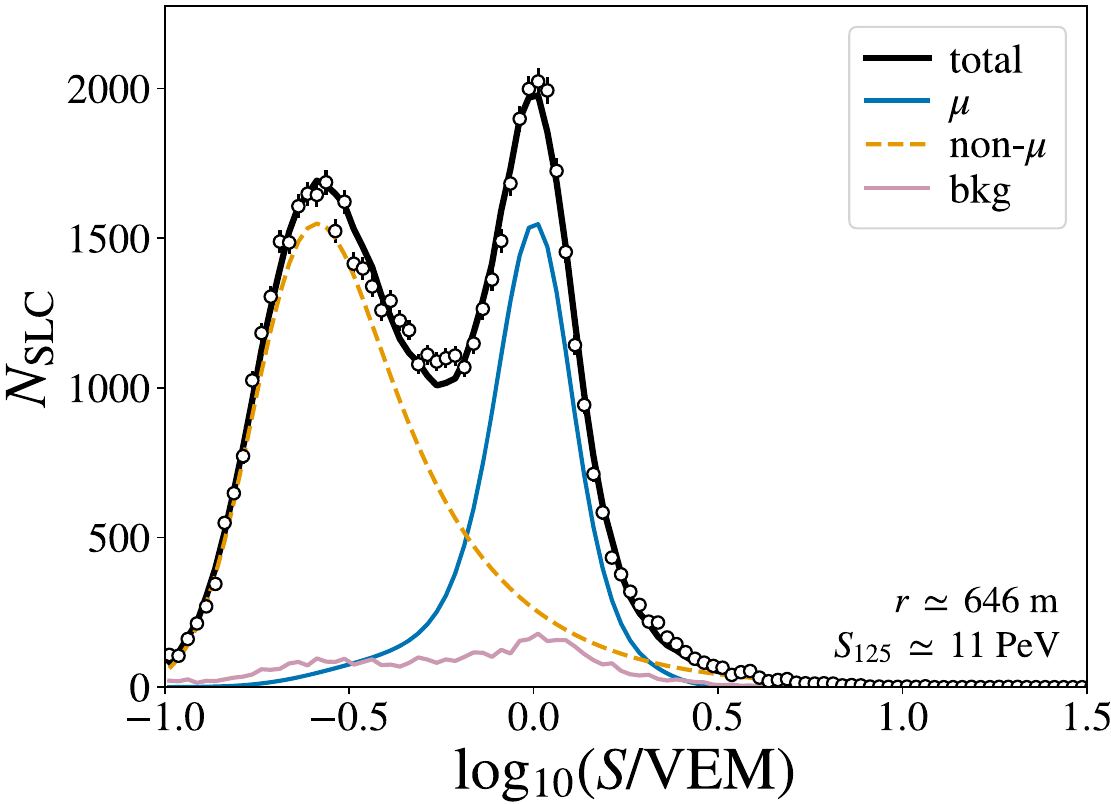}}
    \vspace{-0.4em}
    
    \caption{Signal distribution of tanks at mean lateral distances of 646\,m from the shower axis \cite{IceCubeCollaboration:2022tla}. The distribution is fitted by a model containing the muon contribution and backgrounds. The given $S_{\!125}$ value corresponds to primary energies near 10\,PeV.}
    \label{fig:muon-peak-figure02b}
    \vspace{-1em}
    
\end{figure}

In addition to the high-energy measured in IceCube, with IceTop the dominantly low-energy muons at the surface (\emph{GeV-muons}) can be analyzed and compared to model predictions. Although the surface array has no specific muon detectors, muons can be identified by their energy deposits in the tanks yielding the characteristic peak of minimum-ionizing particles. With increasing distance from the shower axis, this muon peak becomes increasingly prominent in the tank's signal distribution~\cite{IceCubeCollaboration:2022tla}. An example is shown in \cref{fig:muon-peak-figure02b} for a lateral distance of about 650\,m  and a primary energy around 10\,PeV. These signal distributions are fit at fixed energy, zenith, and lateral distance. The model includes individual signal models for the detector response to muons, the electromagnetic part of the air shower, and the contribution from accidental coincident background hits, as shown in the figure. This analysis has been developed in Tom Gaisser's group at the Bartol Research Institute and a detailed description can be found in Ref.~\cite{IceCubeCollaboration:2022tla}.

\begin{figure}[tb]
    \mbox{\hspace{-.5em}%
    \includegraphics[width=0.48\textwidth]{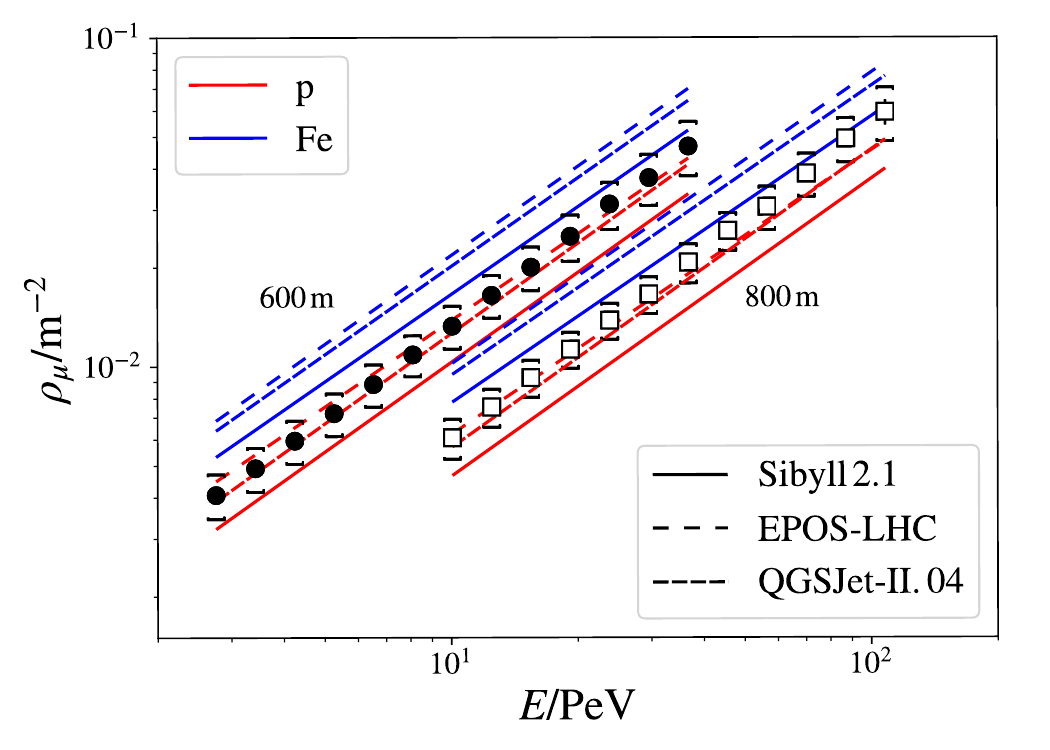}}
    \vspace{-.5em}
    
    \caption{Measured muon density at 600\,m (solid circles) and 800\,m (white squares) lateral distance \cite{IceCubeCollaboration:2022tla}. Error bars indicate the statistical (mostly smaller than the markers), brackets the systematic uncertainty. Also shown are the simulated densities for proton and iron.}
    \label{fig:muon-densities}
    \vspace{-1.em}
    
\end{figure}

\Cref{fig:muon-densities} shows the muon densities derived from three years of IceTop data at distances of $600\,\rm{m}$  and $800\,\rm{m}$ from the shower core, for primary energies between $2.5\,\rm{PeV}$ and $40\,\rm{PeV}$ and between $9\,\rm{PeV}$ and $120\,\rm{PeV}$, respectively. The data are compared to the simulated muon densities obtained from three different hadronic models for proton and iron primaries. The plot shows that the individual models yield different primary compositions. 
This aspect will be further discussed in the following section.

\subsubsection{TeV-muons}

The \emph{TeV-muon} content in air showers can be determined from the energy losses measured in the ice, similar to the analysis of the cosmic-ray mass composition described in \cref{sec:theRoleofMuons}. Therefore, the energy loss of the muons in a bundle is estimated in segments along the bundle's reconstructed trajectory in the ice and is used as an input to a recurrent neural network
layer~\cite{Verpoest:2023qmq}. The output of this layer is combined with $S_{\!125}$ and the reconstructed zenith angle as an input to a number of dense layers which return the air-shower energy and the multiplicity of TeV-muons, $N_\mu$. Further details of this analysis are described in Ref.~\cite{Verpoest:2023qmq}.

\begin{figure*}[tb]
    \vspace{-1.5em}
    \mbox{\hspace{-2.em}\includegraphics[width=0.51\textwidth]{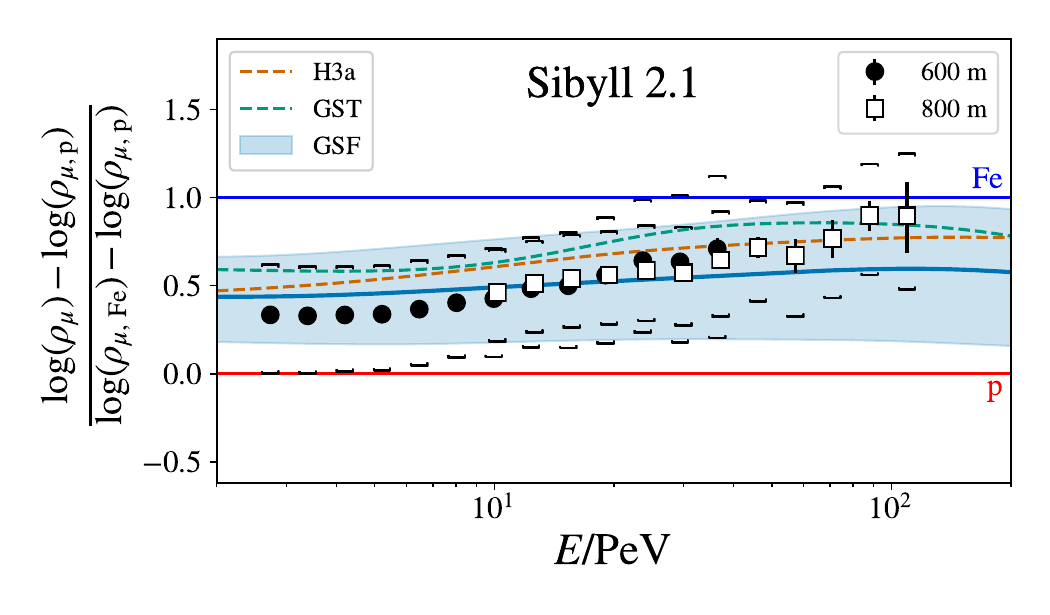}
   \includegraphics[width=0.51\textwidth]{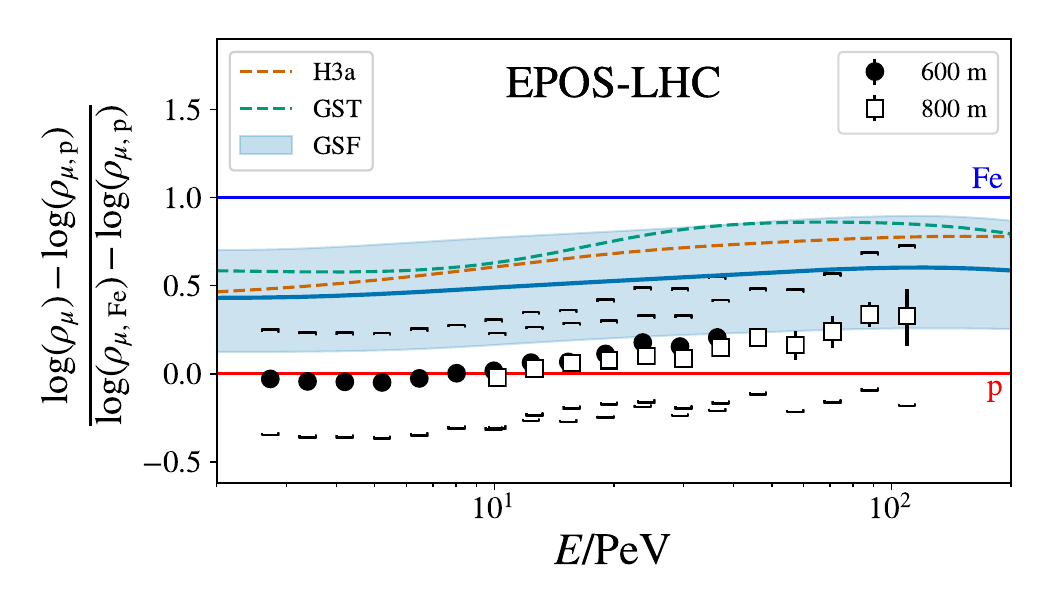}}
    \vspace{-1.em}
    
    \mbox{\hspace{-1.8em}
    \includegraphics[width=0.5035\textwidth]{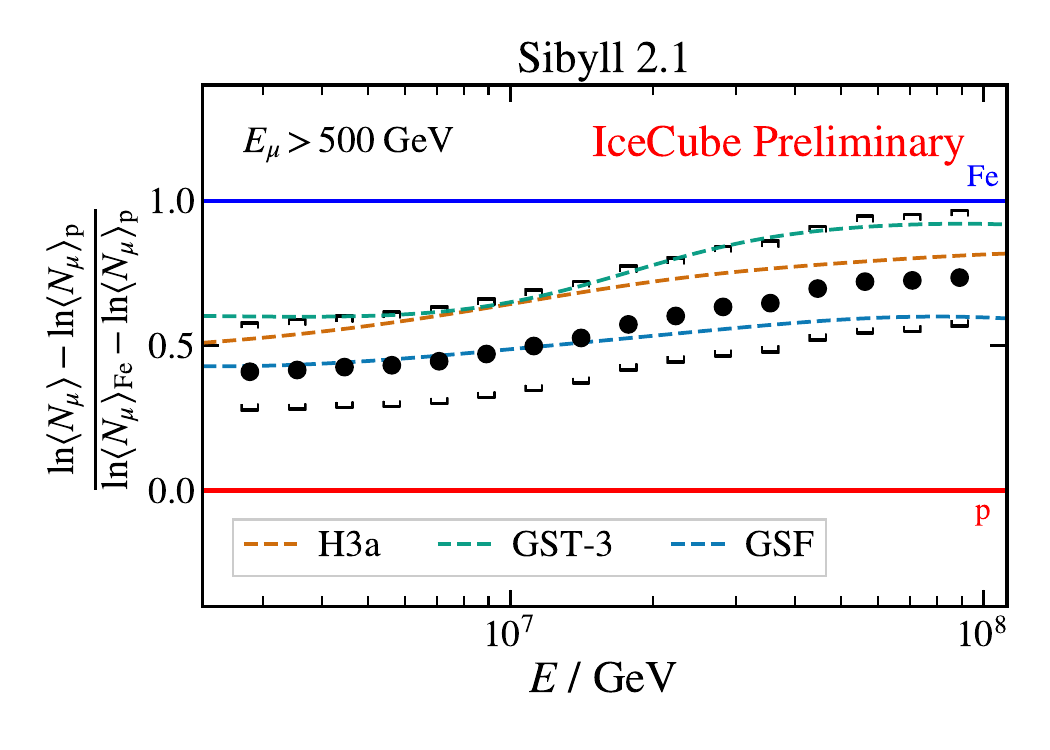}\,\,
    \includegraphics[width=0.5035\textwidth]{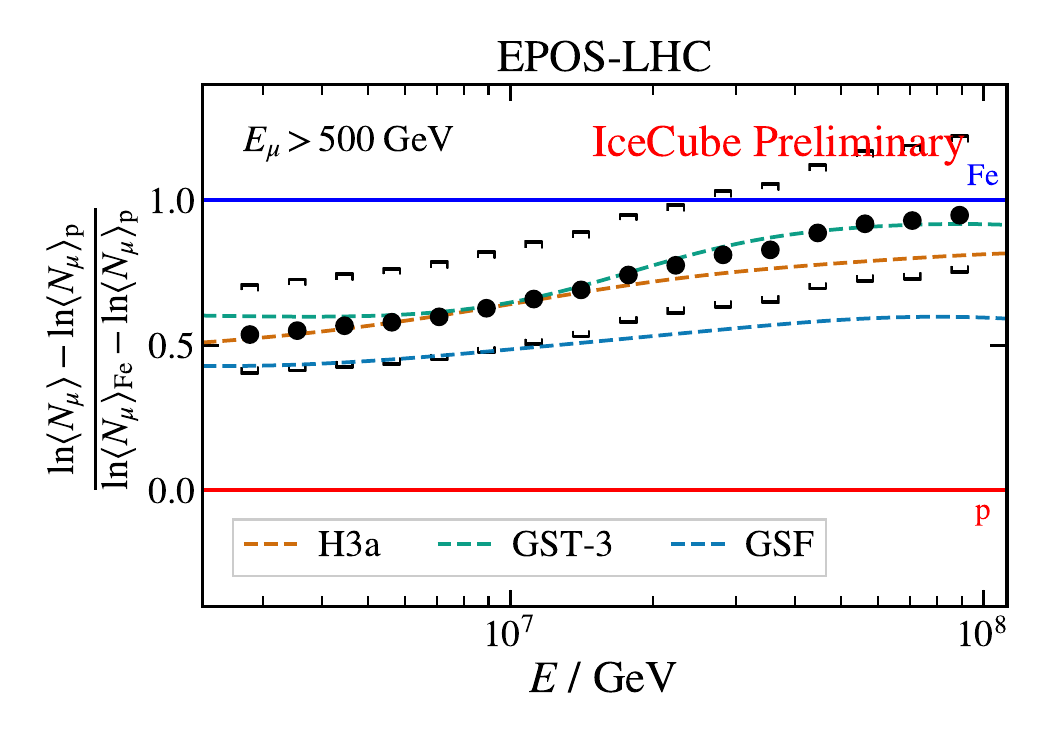}}
    \vspace{-1.5em}
    \caption{Resulting $z$-values from \cref{eq:z} for the GeV-muon density at $600$ and $800\,\rm{m}$ (top) \cite{IceCubeCollaboration:2022tla} and the multiplicity of TeV-muons (bottom) \cite{Verpoest:2023qmq} as function of the primary energy for the hadronic models Sibyll~2.1 (left) and EPOS-LHC (right). The error bars show the statistical uncertainty, while the brackets represent the systematic uncertainties. Also shown are the predictions from the flux models H3a~\cite{Gaisser:2011klf}, GST~\cite{Gaisser:2013bla}, and GSF~\cite{Dembinski:2017zsh}.}
    \label{fig:Steff-ICRC21-sibyll-epos}
\end{figure*}

Using this measurement of the muon multiplicity in the deep ice, the hadronic interaction models Sibyll 2.1, QGSJet-II.04, and EPOS-LHC (the latter two are post-LHC models) have been tested by comparing data to simulations of proton and iron showers \cite{IceCube:2021ixw,Verpoest:2023qmq}. 

To compare the measured muon densities to predictions from different hadronic interaction models with certain cosmic-ray flux assumptions in more detail, one defines the quantity
\begin{equation}\label{eq:z}
    z = \frac{\ln\langle x_{\mathrm{data}}\rangle - \ln\langle x_\mathrm{p}\rangle}{\ln\langle x_\mathrm{Fe}\rangle - \ln\langle x_\mathrm{p}\rangle} \, ,
\end{equation}
where $x$ is the muon content in air showers, e.g., $N_\mu$ or $\rho_\mu$ obtained from data. The quantities $x_\mathrm{p}$ and $x_\mathrm{Fe}$ are derived from proton and iron simulations, respectively, employing one of the hadronic models. Hence, the resulting $z$-values are model-dependent.

\Cref{fig:Steff-ICRC21-sibyll-epos} shows the $z$ distributions for primary energies between 2.5 PeV and 100 PeV for the TeV-muon content in air showers obtained from data \cite{Verpoest:2023qmq}, $N_\mu$ (bottom), compared to the measurement of the densities of GeV-muons at 600 and 800\,m, $\rho_\mu$ (top), described in \cref{subsec:GeV_muons}. The distributions are shown for the hadronic interaction models Sibyll 2.1 and EPOS-LHC. If the models give a realistic description of experimental data, for GeV- and TeV-muons the $z$-values at a given energy should be the same. 

However, while the distributions based on Sibyll 2.1 show a reasonable agreement, internal inconsistencies can be observed within EPOS-LHC (and other post-LHC models~\cite{Verpoest:2023qmq,IceCubeCollaboration:2022tla}). This is because post-LHC models predict more GeV-muons than Sibyll 2.1 over the entire energy range, yielding a very light mass composition which is not consistent with the measurement described in \cref{subsec:TeV_muons} or with any current flux model that is consistent with experimental data~\cite{Gaisser:2011klf,Gaisser:2013bla,Dembinski:2017zsh}. The predicted multiplicities of TeV-muons, however, agree between the models within uncertainties.

In air showers initiated by cosmic rays above about 1\,EeV, the measured densities of low-energy muons (around 1\,GeV) are always higher than the predictions from simulations, even taking non-physical values above the predictions for iron~\cite{PierreAuger:2014ucz,PierreAuger:2016nfk,Dembinski:2019uta,Soldin:2021wyv}. This problem, referred to as the \emph{Muon Puzzle}, could not be solved by tuning the models with LHC data~\cite{Albrecht:2021cxw}. However, in IceCube data, at least up to 100\,PeV, there is no indication that the muon densities could be too high. At the lowest primary energies the post-LHC models rather tend to predict high muon densities. As discussed in Refs.~\cite{Soldin:2021wyv,Albrecht:2021cxw}, it appears that the disagreement between realistic predictions and observed muon densities increases with increasing energy.

These inconsistencies render it impossible to unambiguously determine the mass composition of cosmic rays by employing muon multiplicities. Therefore, it is of prime importance to further investigate the causes for the inconsistencies and to improve the models. Improved analysis methods are under development~\cite{IceCube:2023lhg,IceCube:2023suf} which will allow for simultaneous measurements of GeV- and TeV-muons in the same air shower on an event-by-event basis and thereby provide additional constraints for hadronic models~\cite{Albrecht:2021cxw}.

\subsection{High-energy muon spectrum} \label{sec:HE_muon_flux}
High-energy muons are produced early during the development of air showers, mainly from the decay of pions and kaons. However, at very high muon energies, above $1\,\mathrm{PeV}$, the prompt contribution from leptonic decays of short-lived heavy hadrons and unflavored vector mesons is expected to dominate the total flux of atmospheric muons \cite{Enberg:2008te,Fedynitch:2018cbl}. In IceCube, these high-energy muons are generally accompanied by a bundle of lower-energy muons above threshold ($>460\,\mathrm{GeV}$), which forms the most compact region of the shower core. As shown in Ref.~\cite{IceCube:2015wro}, any bundle muon with an energy above about $30\,\mathrm{TeV}$ will likely be the most energetic muon in the bundle, the \textit{leading muon}. 

\begin{figure}[tb]
    \mbox{\hspace{-0.5cm}\includegraphics[width=0.513\textwidth]{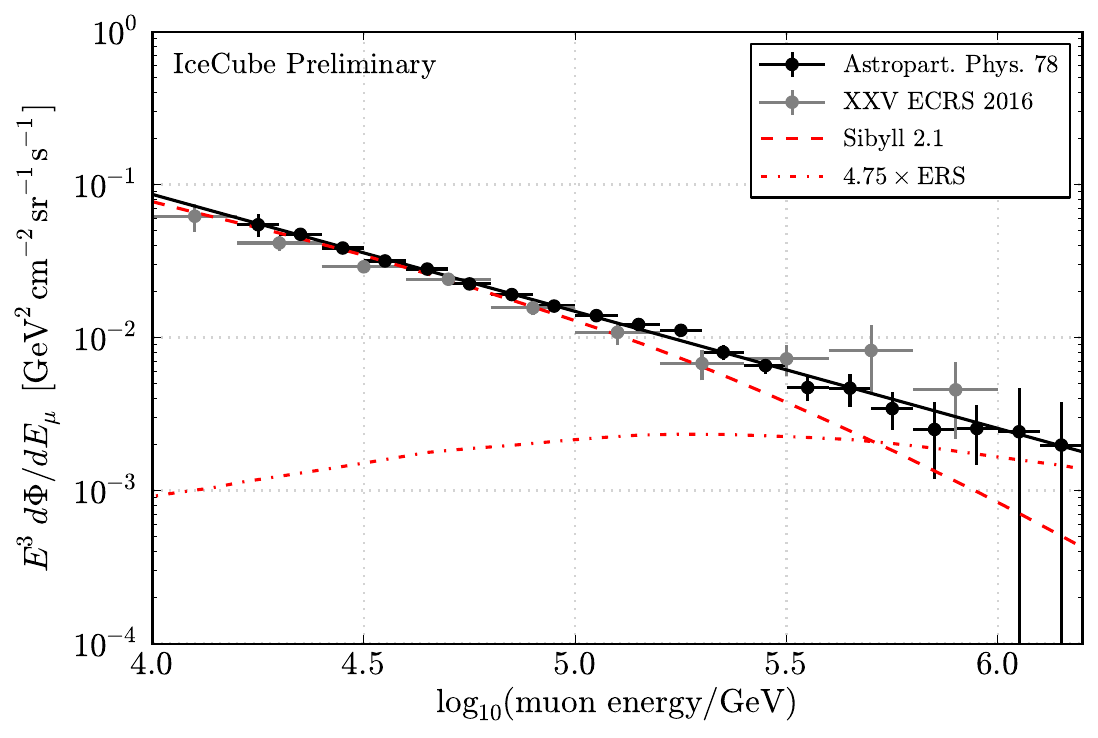}}
    \vspace{-2em}
    \caption{High-energy muon energy spectra measured in IceCube, taken from Ref. \cite{IceCube:2015wro} (cut-based approach) and Ref. \cite{Fuchs:2017nuo} (machine learning approach). The solid line represents the power law fit from \cref{eq:HE_spec}. Also shown are predictions of the conventional muon flux, using Sibyll 2.1 as hadronic model and assuming a H3a primary flux, as well as the best fit ERS prompt flux \cite{Enberg:2008te} (see text for details).}
    \label{fig:HE_spectrum}
    \vspace{-0.5em}
\end{figure}

The selection and the energy reconstruction of this most energetic muon is generally based on the energy loss characteristics in the ice. Beyond the critical energy of around $1\,\rm{TeV}$, the muon energy loss in the ice is highly dominated by stochastic (radiative) energy losses. As described in Ref.~\cite{IceCube:2015wro}, the energy loss profile can therefore be used to select the most energetic muon and to estimate its energy at the surface in order to derive the muon spectrum above $10\,\mathrm{TeV}$. This has been done using two independent approaches: a cut-based event selection using two years of IceCube data \cite{IceCube:2015wro}, which was developed in Tom Gaisser's group at the Bartol Institute, and a machine learning approach based on one year of data \cite{Fuchs:2017nuo}.

\Cref{fig:HE_spectrum} shows the resulting muon energy spectra at surface level (about $690\,\mathrm{g/cm^2}$). Within the accuracy of these analyses, the spectrum can be  approximated by a simple power law of the form
\begin{equation}
    \frac{{\rm d}\Phi}{{\rm d}E_\mu}= \frac{0.86 \times 10^{-10}}{\mathrm{TeV\,cm^{2}\,sr\,s}} \left( \frac{E_\mu}{10\,\mathrm{TeV}} \right)^{-3.76} .
    \label{eq:HE_spec}
\end{equation}
Simulated Monte Carlo predictions are also shown in \cref{fig:HE_spectrum}~\cite{Soldin:2018vak}, using Sibyll 2.1 \cite{Ahn:2009wx} as hadronic interaction model and the H3a cosmic-ray flux assumption from Ref.~\cite{Gaisser:2011klf}. Monte Carlo predictions underestimate the experimental data towards high energies, which is expected to be caused by a missing prompt muon component in Sibyll 2.1. As described in Ref.~\cite{IceCube:2015wro}, this missing component is fit with multiples of the prompt ERS flux $\Phi_\mathrm{ERS}$ \cite{Enberg:2008te}. Assuming an H3a primary flux, the best fit yields a prompt flux of \mbox{$\Phi_\mathrm{prompt}=4.75\times \Phi_\mathrm{ERS}$}. This estimate strongly depends on the underlying primary flux and the corresponding systematic uncertainties are therefore very large, ranging from \mbox{$\Phi_\mathrm{prompt}=0.94\times \Phi_\mathrm{ERS}$} to \mbox{$\Phi_\mathrm{prompt}=6.97\times \Phi_\mathrm{ERS}$} (see Ref. \cite{IceCube:2015wro} for further details). However, the non-existence of a prompt muon flux can not be excluded yet, with statistical significances from $1.52\,\sigma$ up to $5.24\,\sigma$, depending on the primary flux assumption (see Ref.~\cite{IceCube:2015wro} for details).

\begin{figure}[tb]
    \mbox{\hspace{-0.4cm}\includegraphics[width=0.52\textwidth]{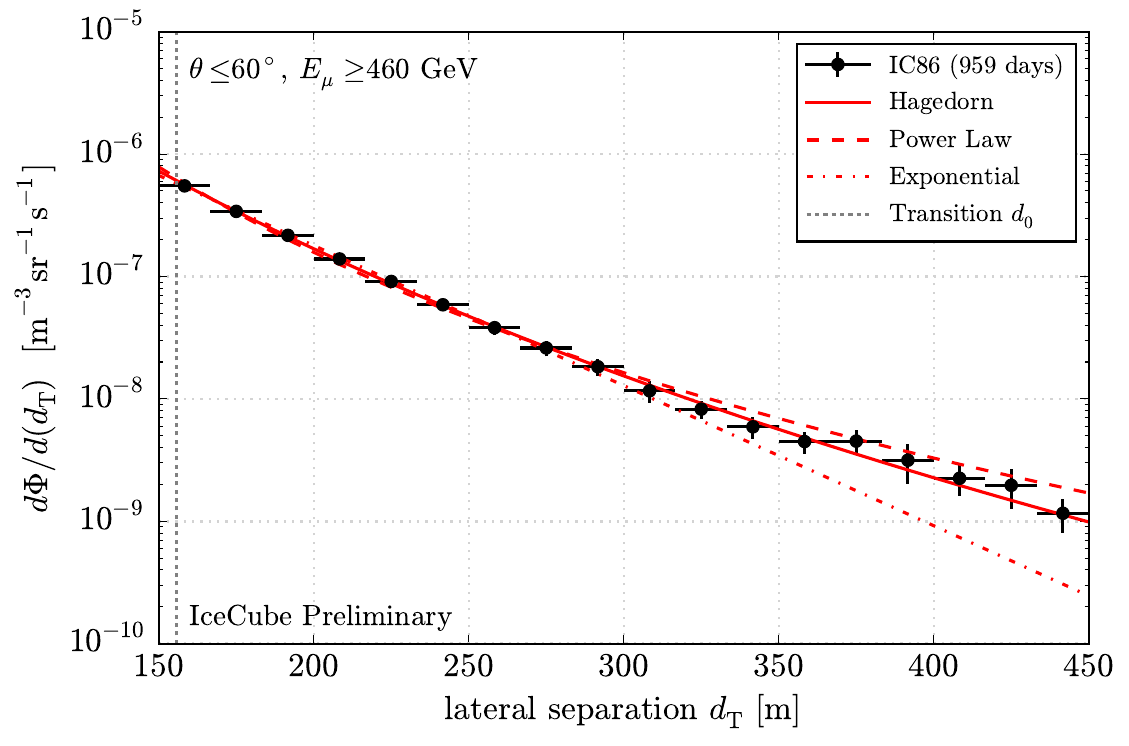}}
    \vspace{-2em}
    \caption{Differential flux of laterally separated muons with energy above $460\,\mathrm{GeV}$ and zenith angle $\theta\leq 60^\circ$, obtained from three years of IceCube data~\cite{Soldin:2018vak}. Also shown is the corresponding Hagedorn fit of the form of \cref{eq:hagedorn}, as well as an exponential and a power law fit for comparison (see text for details).}
    \label{fig:LS_dT1}
    \vspace{-0.5em}
\end{figure}

\begin{figure*}[tb]
    \centering
    \includegraphics[width=1.00\textwidth]{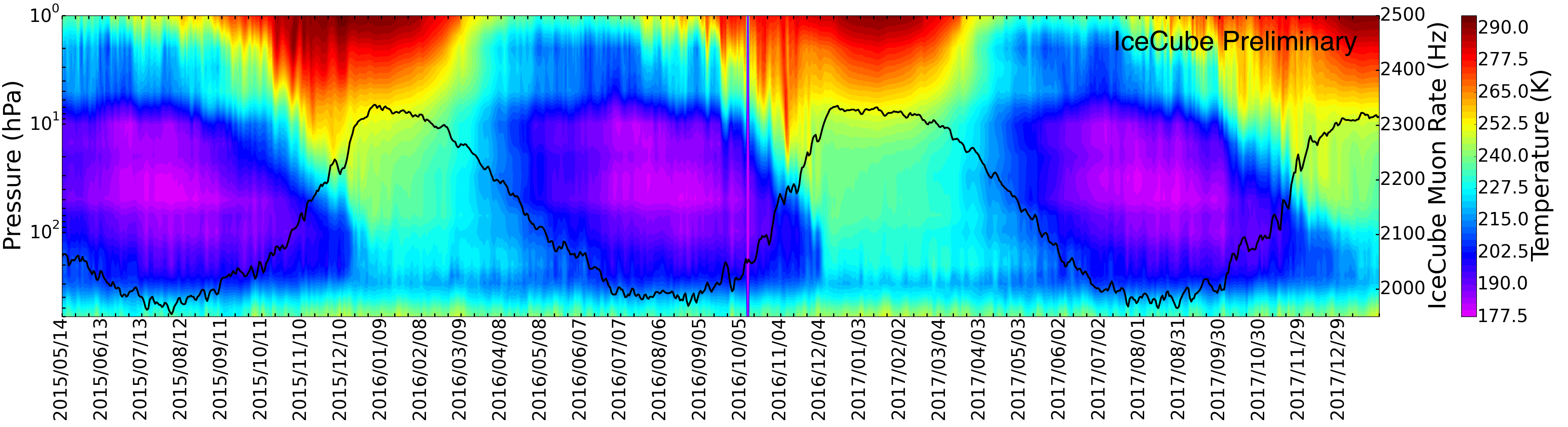}
    \vspace{.1em}
    
    \hspace{-3em}\includegraphics[width=.92\textwidth]{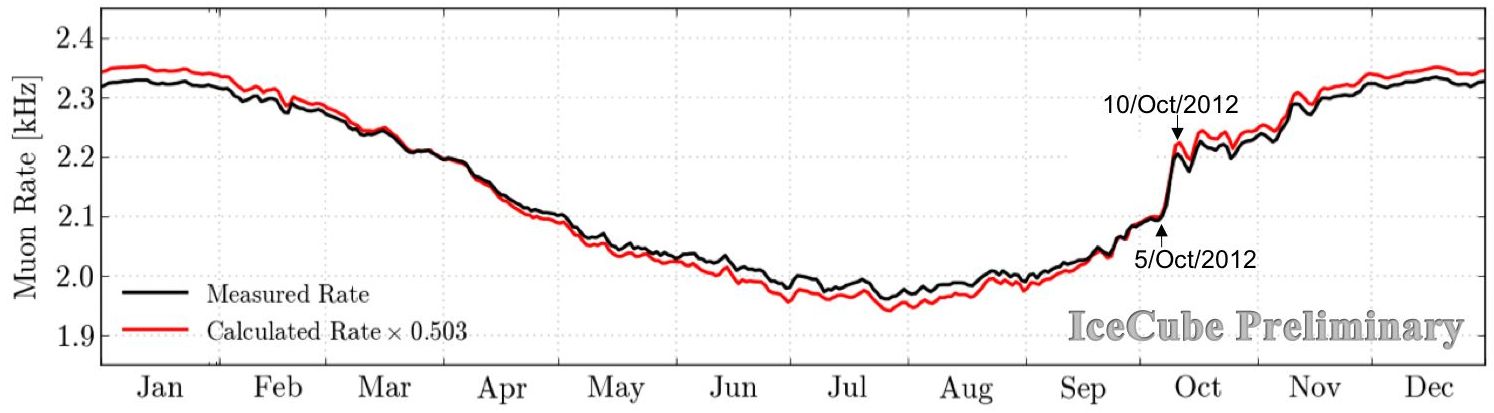}
    \caption{\textit{Top:} IceCube muon rate (black line) overlaid with the temperature profile of the South Pole atmosphere at different pressure heights~\cite{Tilav:2019xmf}. The plot illustrates the behavior of the seasonal cycles as well as the short-term (day to week time scales) variations in rate with respect to the temperature variations in the stratosphere. \textit{Bottom:} Comparison of the measured with the calculated muon rate for 2012 data from Ref.~\cite{Tilav:2019xmf}.}
    \label{fig:muon-seasonal-Fig4-ShortTermDetails}
\end{figure*}

\subsection{Laterally separated muons}
\label{sec:LS_munos}
One expects that hadrons with large transverse momentum, $p_\mathrm{T}\gtrsim 2\,\mathrm{GeV/c}$, are produced in high-energy air showers which can subsequently decay into muons. These muons separate from the shower core while traveling to the ground, forming laterally separated muons with distances up to several $100\,\mathrm{m}$ from the dense core region. The resulting lateral separation is a direct measure of the $p_\mathrm{T}$ of the parent hadron. Experimentally a transition from soft to hard interactions is observed in the $p_\mathrm{T}$ spectrum, which falls off exponentially with a transition to a power law at approximately $2\,\mathrm{GeV/c}$, where interactions can be described in the context of pQCD~\cite{Hagedorn:1983wk}. This transition should be also visible in the lateral separation distribution of muons.

The bright muon bundle together with the isolated, laterally separated muon form a distinct double-track signature in IceCube~\cite{IceCube:2012ldb}. The lateral distance of the laterally separated muon to the shower core is approximately given by
\begin{equation}
    d_\mathrm{T}\simeq \frac{p_\mathrm{T}\, H}{E_\mu\, \cos(\theta)}\, ,
    \label{eq:dT}
\end{equation}
where $p_\mathrm{T}$ is the transverse momentum of the muon, $E_\mu$ is the muon energy, $\theta$ is the zenith angle direction, and $H$ is the altitude of hadron production. In IceCube, the muon bundle in the central core region and the isolated, laterally separated muon are reconstructed simultaneously to obtain the lateral separation. \Cref{fig:LS_dT1} shows the flux of laterally separated muons on surface level, obtained from three years of IceCube data~\cite{Soldin:2018vak}. In order to derive the flux at surface level, effective areas obtained from simulations are used. The event selection is based on a previous analysis, which used one year of data from IceCube in its 59-string configuration and is described in Ref.~\cite{IceCube:2012ldb}. Also shown in \cref{fig:LS_dT1} is a QCD-inspired \emph{Hagedorn fit} \cite{Hagedorn:1983wk} of the form
\begin{equation}
    \frac{{\rm d}\Phi}{{\rm d}\,d_\mathrm{T}}= \frac{236.3}{{\rm m}^3\,{\rm sr\,s}}\, \left( 1+\frac{d_\mathrm{T}}{d_0} \right)^{-9.7} ,
    \label{eq:hagedorn}
\end{equation}
with \mbox{$d_0=(157.3\pm 43.0)\,\mathrm{m}$}. This functional form behaves like an exponential for \mbox{$d_\mathrm{T}/d_0\rightarrow 0$} and describes a power law for \mbox{$d_\mathrm{T}/d_0\rightarrow \infty$}, with the transition around $d_0$. Also shown are fits assuming a pure exponential and a simple power law. The Hagedorn function describes the experimental distribution well (\mbox{$\chi^2/\mathrm{ndof}=20.16/16$}), with the transition from soft to hard interactions at around $157.3\,\mathrm{m}$. In contrast, the pure exponential and power law fits are in poor agreement with the data, especially towards large separations (\mbox{$\chi^2/\mathrm{ndof}=97.92/16$} and \mbox{$\chi^2/\mathrm{ndof}=53.60/16$}, respectively). Within uncertainties, the measured flux, as well as the resulting fit parameters, are in agreement with previous results \cite{IceCube:2012ldb}. Comparison with model predictions in this analysis is limited to the hadronic interaction model Sibyll 2.1 which agrees with the experimental data within the uncertainties~\cite{Soldin:2018vak}.

\subsection{Seasonal variations of atmospheric lepton rates} \label{sec:SeasonalVariationsMuonNeutrinoRates}

\subsubsection{Muon rates} 

The rate of high-energy muons in the deep ice, which are produced in the first interactions of cosmic rays with energies in the TeV range and above, exhibits seasonal variations which are correlated with the temperature. In the energy range where the interaction lengths of the muons' parent mesons are comparable to their decay lengths, higher temperatures lead to lower density, and hence to relatively more decays. This in turn leads to to higher muon production rates \cite{Gaisser:2021cqh,Seasonal}. Data from IceCube taken between May 2015 and December 2017 are shown in \cref{fig:muon-seasonal-Fig4-ShortTermDetails} (top). Temperature variations cause variations in the density which then change the interaction probability of particles with the atmosphere. The interactions are in competition with decays -- less interaction results in more decays with muons and neutrinos as decay products. The variation of the rate $R$ is characterized by a correlation coefficient $\alpha_{T}(E_\mu)$ which is employed to describe the rate change as a linear function of the change of the so-called effective temperature:
\begin{equation}
     \frac{\delta R}{\langle R\rangle} = \alpha_{T}\, \frac{\delta T_{\rm eff}}{\langle T_{\rm eff}\rangle}\, .
    \label{eq:alpha_T}
\end{equation}
The effective temperature $T_{\rm eff}(\theta)$ is defined by the measured temperature profile (at a zenith angle $\theta$) folded with the muon production spectrum and effective area for muon detection, integrated over the muon energy.

Since muons mainly come from decays of pions and kaons, which have different lifetimes and interaction probabilities, the coefficient $\alpha_{T}$ is sensitive to the relative contributions of pions and kaons. Therefore the measurement of $\alpha_{T}$ yields another input for tuning hadronic models. The comparison of muon rate data for a specific year with the corresponding calculations in \cref{fig:muon-seasonal-Fig4-ShortTermDetails} (bottom) shows a remarkable agreement (see also Ref.~\cite{Tilav:2019xmf}). 
Features, even small ones, are well reproduced by analytical calculations which were one of the major research interests of Tom Gaisser throughout the last decade. The correction factor $0.503$ accounts for a known failure of simulations to predict the absolute rate in IceCube at detector level which affects the effective area used in the calculations. In addition, data show a somewhat higher $\Delta R$ amplitude yielding \mbox{$\alpha_{T}^{\rm{meas}}=0.75$} compared to the calculated rate \mbox{$\alpha_{T}^{\rm{calc}}=0.85$}. Thus, Tom Gaisser also noticed that calculations can still be refined, e.g., by using a temperature profile instead of averages, account for the muon multiplicity in the bundle, and the mass composition. This was one of the pending projects for Tom Gaisser which he could not finalize (for further discussion, see also Refs.~\cite{Gaisser:2021cqh,Seasonal}). 

\begin{figure}[tb]
    \vspace{0.4em}
    \mbox{\hspace{-1.2em}
    \includegraphics[width=0.52\textwidth]{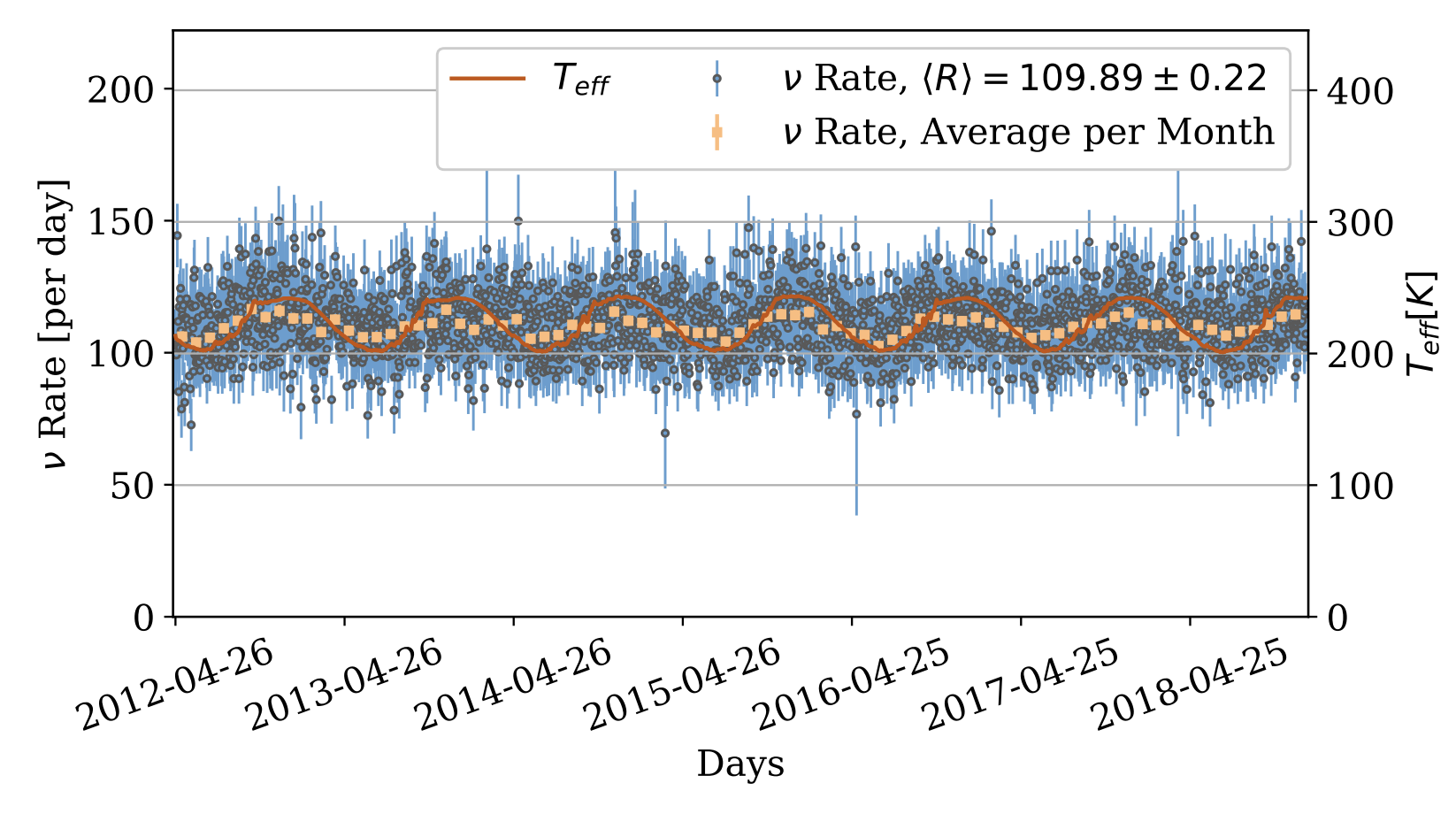}}
    \vspace{-2em}
    \caption{Seasonal variations of the atmospheric muon neutrino flux in IceCube~\cite{Abbasi:2023cun}. The neutrino data is shown in daily  (blue points/error bars) and monthly bins (orange points/error bars). The red line depicts the calculated effective temperature.}
    \label{fig:neutrino_seasonal}
    \vspace{-0.8em}
\end{figure}

\subsubsection{Neutrino rates} 
A similar study has been done for seasonal variations of atmospheric neutrino rates \cite{Abbasi:2023cun,IceCube:2023ezh}. While the muon data are obtained for the Southern hemisphere the neutrino data contain complementary information from the Northern hemisphere, though with lower statistical significance. It adds to the complementarity that a neutrino has another kinematic relation with its parent particle.

\Cref{fig:neutrino_seasonal} shows the atmospheric muon neutrino flux measured by IceCube between April 2012 and April 2018 with an average rate of about 110 neutrinos per day (260,000 total events). A significant (\mbox{$>\!10\,\sigma$}) correlation between the neutrino rate and the atmospheric temperatures of the stratosphere is observed. For the observed $10\%$ seasonal change of
effective atmospheric temperature, a $3.5\%$
change in the muon neutrino flux is reported. However, this deviates from the expected correlation of $4.2\%$ as obtained from theoretical predictions (assuming Sibyll 2.3, see Ref.~\cite{Abbasi:2023cun} for details). In fact, considering the systematic and statistical uncertainties, the results are in tension with predictions at the \mbox{$1-2\,\sigma$} level. Despite careful studies of the systematic uncertainties this tension is currently not understood.

\subsection{IceCube cosmic-ray anisotropy} \label{sec:CR_Anisotropy}

\subsubsection{Energy dependence of the cosmic-ray anisotropy}\label{subsec:CR_Anisotropy_energy_dep}

\begin{figure}[tb]
\vspace{0.8em}
    \mbox{\hspace{-1.5em}
    \includegraphics[width=0.54\textwidth]{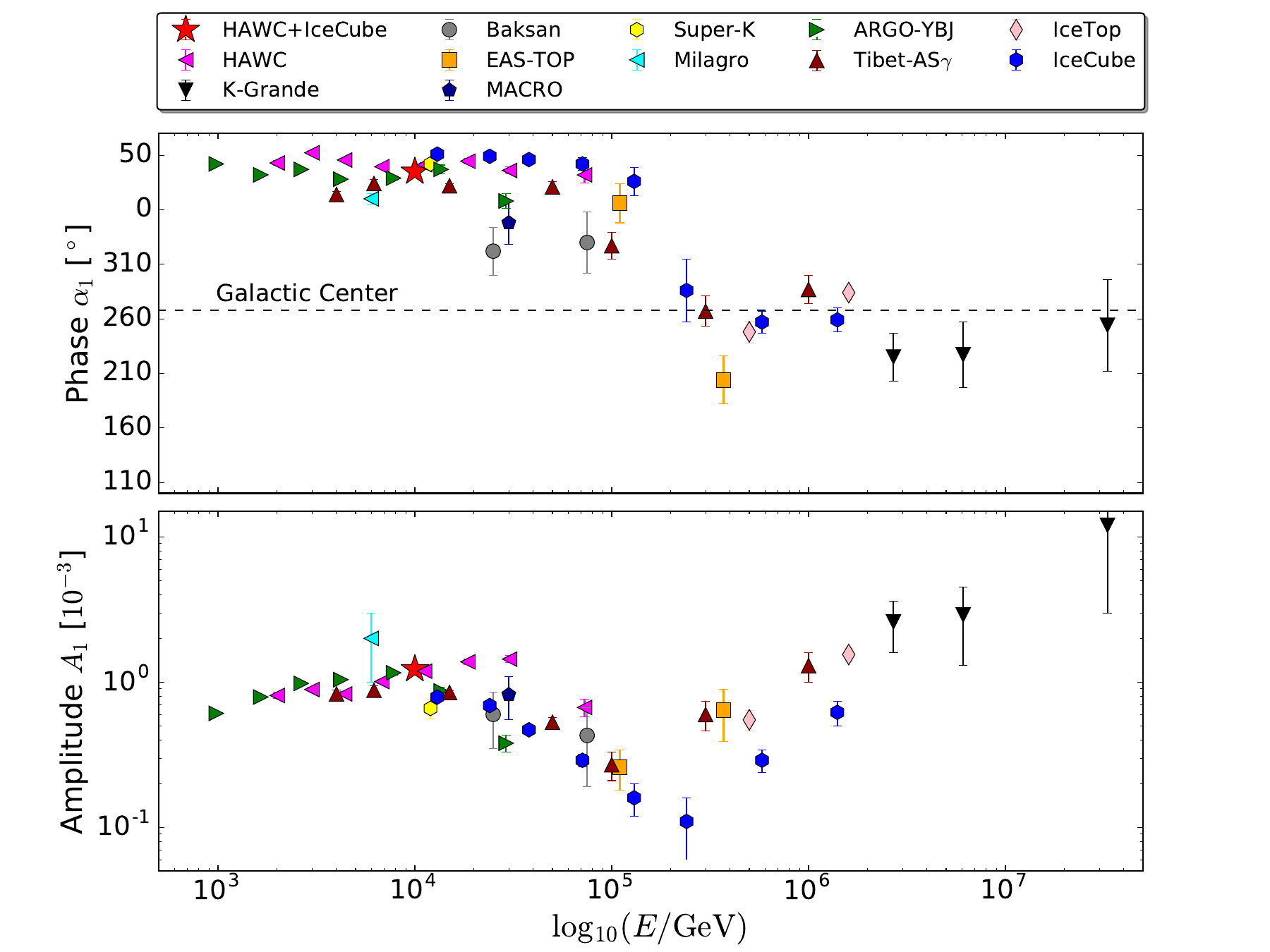}}
    \caption{Reconstructed dipole phase (top) and amplitude (bottom) from published TeV--PeV results from various experiments. For details see Ref.~\cite{Ahlers:2016rox}.}
    \label{fig:CRAI3HW2yr_faseyampl}
    \vspace{.5em}
\end{figure}

\begin{figure*}[tb]
    \vspace{-1.5em}
    \mbox{\hspace{0em}
    \includegraphics[width=0.48\textwidth]{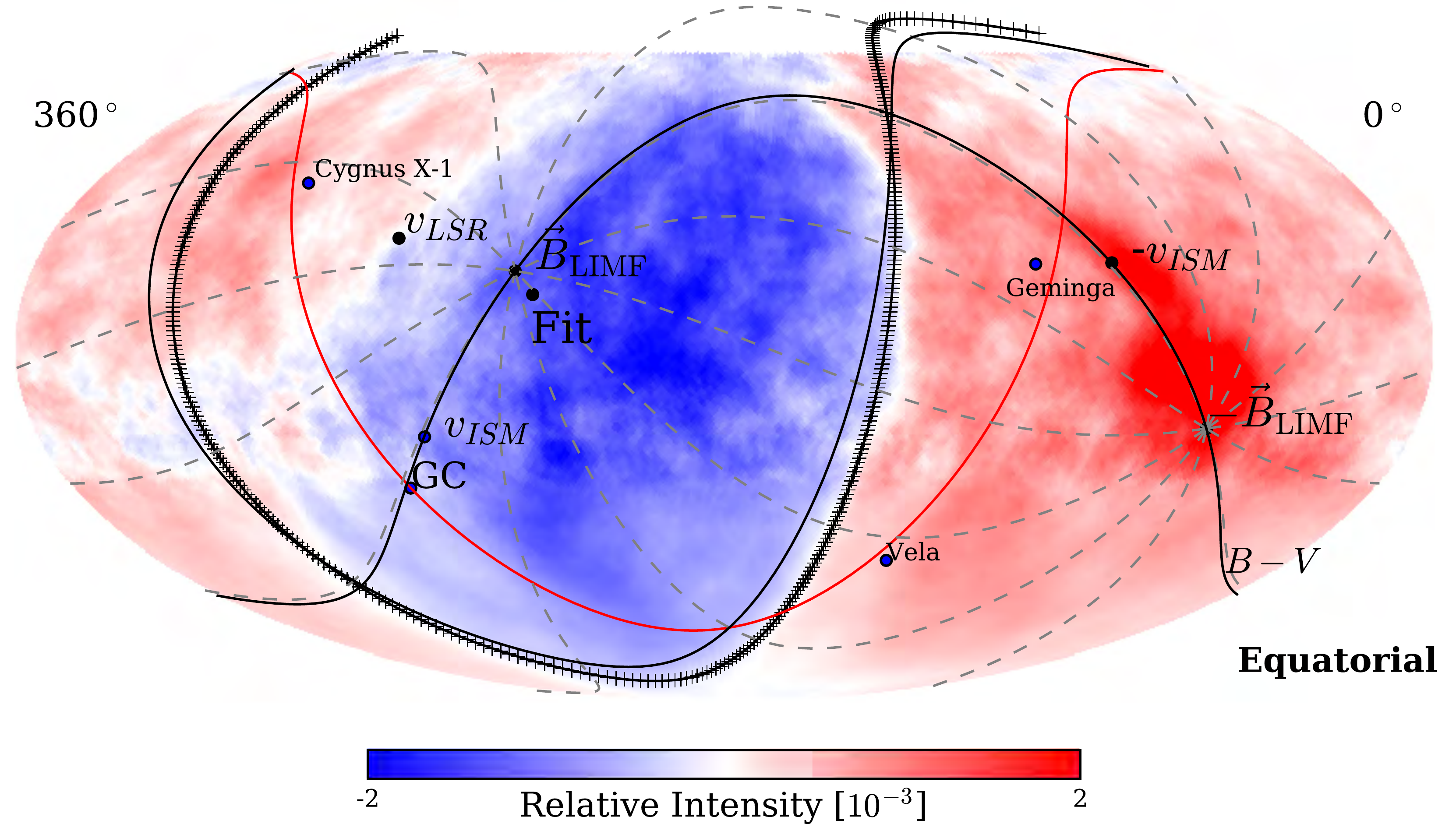}\;\;\;\;\includegraphics[width=0.48\textwidth]{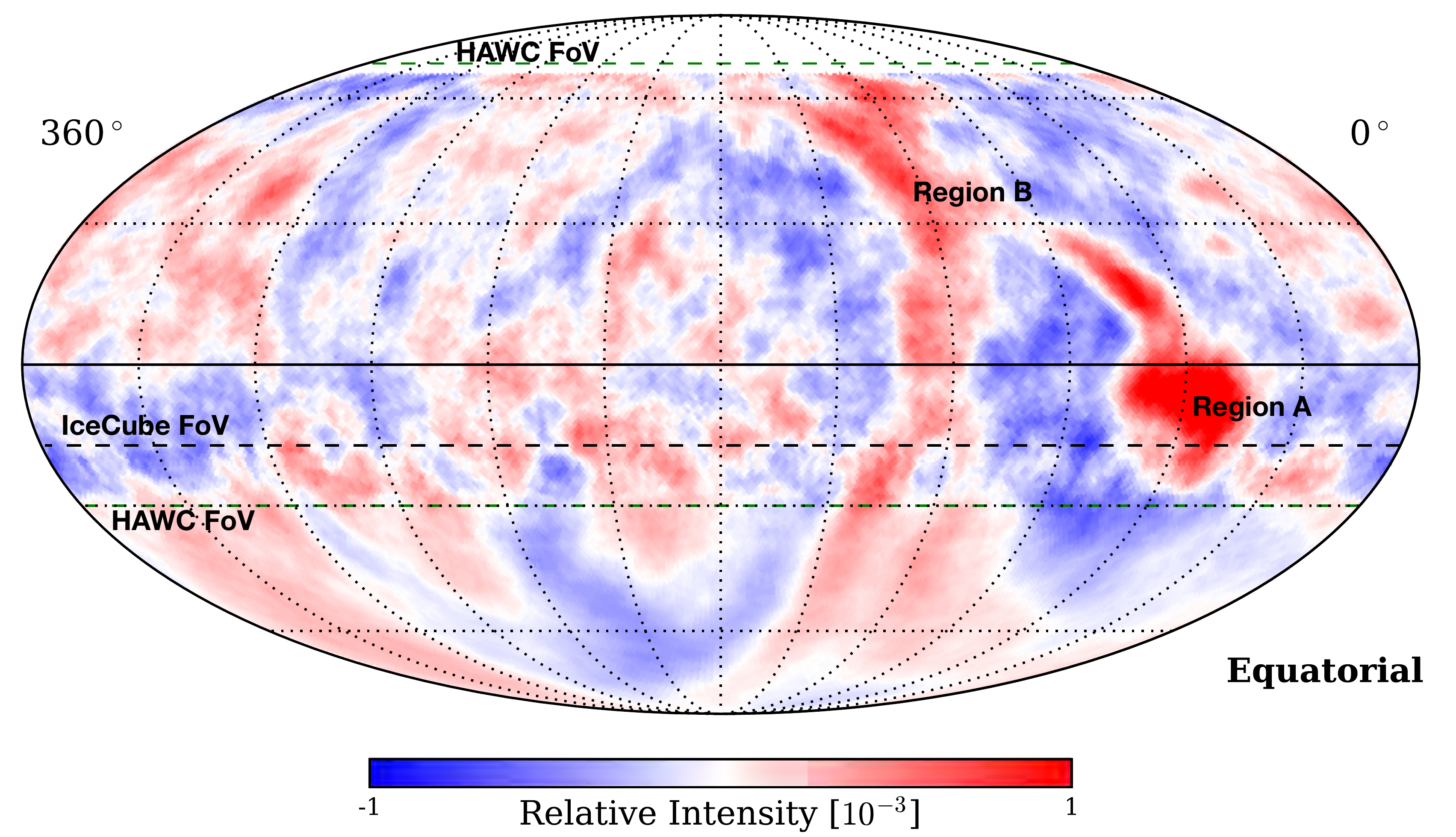}
    }
    \caption{\emph{Left:} Relative intensity of cosmic rays at 10\,TeV median energy in J2000 equatorial coordinates. The fit to the boundary between large scale excess and deficit regions is shown as  black crosses. The  black curves are the magnetic equator from Ref.~\cite{Zirnstein_2016} and the plane containing the local interstellar medium magnetic field and velocity ($B-V$ plane). The Galactic plane is shown as a red curve. Indicated are also the Sun directions in the local rest frame (LSR) and in the interstellar medium (ISM). \emph{Right:} Relative intensity after subtraction of the fitted leading multipoles~\cite{HAWC:2018wju}.}
    \label{fig:CRAI3HW2yr_LIMF}
\end{figure*}

IceCube has studied the cosmic-ray anisotropy in a wide energy range between 10\,TeV and 5\,PeV \cite{IceCube:2016biq,IceCube:2012vve}. The arrival directions have been derived from muons in the deep ice, except for some measurements in the PeV range where the direction has been obtained from  showers in IceTop. While the shower measurements directly yield an estimate of the primary energy, for muons, the primary energy has been estimated from the energy deposited in the ice. The main features of the observed anisotropy are
\begin{itemize}
\setlength{\itemsep}{3pt}%
    \item a dominant dipole at a relative intensity level of about $10^{-3}$,
    \item a significant small scale structure at a level of $10^{-4}$,
    \item a phase shift of the dipole around 150\,TeV,
    \item a turning point of the dipole amplitude at about 10\,TeV. 
\end{itemize}

If the anisotropies are due to magnetic fields the latter observation could indicate a transition from the heliosphere to the interstellar magnetic field. The energy dependencies of the dipole phase and amplitude are depicted in \cref{fig:CRAI3HW2yr_faseyampl}. Details of effects of magnetic fields require anisotropy analyses over the full sky. This cannot be achieved by a single experiment because of the restricted field of view. For example, IceCube sees the declination band from $-16^\circ$ to $-90^\circ$.%

\subsubsection{IceCube/HAWC  all-sky anisotropy at 10 TeV} \label{sec:IceCube/HAWC_Anisotropy}

A full-sky analysis of the cosmic-ray arrival direction distribution has been performed with data collected by the HAWC and IceCube Observatories in the northern and southern hemispheres, respectively, at the same median primary particle energy of 10\,TeV \cite{HAWC:2018wju}. 

The combined sky map of the relative intensity of cosmic rays, that is the deviation from the intensity average in a declination band, is shown in \cref{fig:CRAI3HW2yr_LIMF}. While the left plot shows clearly the dominance of a dipole, the right panel shows the small-scale structures remaining after subtraction of the leading multipoles with $\ell \leq 3$ in order to reveal structures smaller than $60^\circ$. The multipoles are determined by fitting spherical harmonics $Y_{lm}(\delta, \phi)$ to the sky map ($\delta,\ \phi$ are declination and right ascension in equatorial coordinates). In this fit the \mbox{$m=0$} components, describing North--South effects, cannot be determined because the  relative intensities are taken with respect to the average in a declination band.

\subsubsection{Local interstellar magnetic field and heliosphere} \label{sec:LIMF&Heliosphere}
The combined HAWC--IceCube analysis largely eliminates biases that result from partial sky coverage. The full sky coverage allows us to probe into the propagation of TeV cosmic rays through our local interstellar medium and the interaction between the interstellar and heliospheric magnetic fields. 

Scattering on magnetic turbulences is a diffusive process and would, on large scales, lead to isotropy. Therefore anisotropies are expected to originate from local effects (local sources, locally aligned fields, etc.) or movements, like the Compton-Getting effect due to the movement of the Earth around the Sun. 

\Cref{fig:CRAI3HW2yr_LIMF} (left) shows directional correlations between the anisotropy structures and features of the local interstellar magnetic field (LIMF). An estimate of the dipole direction is obtained by fitting a plane along the boundary between large scale excess and deficit. The fitted dipole axis points roughly into the direction of the LIMF, as determined for distances up to $\sim 1000\,\mathrm{AU}$ from the Sun by independent observations~\cite{Zirnstein_2016}. If one assumes the dipole to be aligned with the LIMF one could estimate the missing  North-South dipole component (\mbox{$m=0$}). A more detailed discussion can be found in Ref.~\cite{HAWC:2018wju}.

\subsection{Measurements of the Moon and Sun shadows } \label{sec:Moon_Sun_Shadow}
\begin{figure}[tb]
    \vspace{-1.2em}
    \centering
    \includegraphics[width=0.46\textwidth]{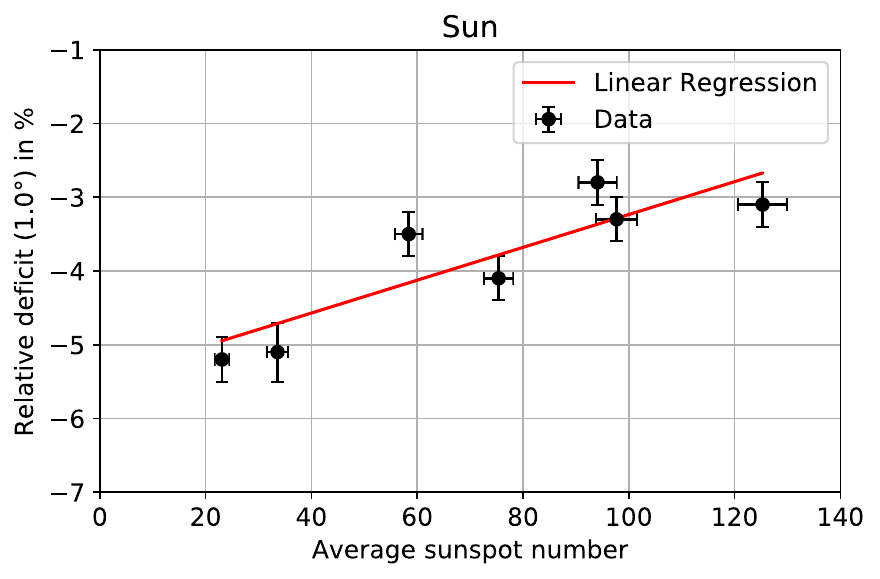}\\
    \includegraphics[width=0.45\textwidth]{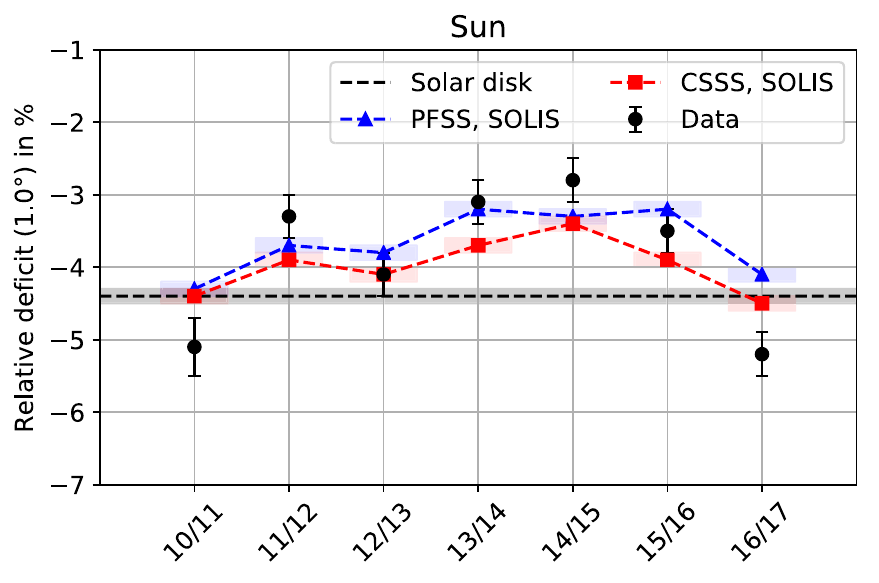}
    \caption{\emph{Top:} Correlation of measured relative deficit due to the Sun shadow and average sunspot number. \emph{Bottom:} Comparison of the measured relative deficit due to the Sun shadow to the deficit expected from different models of the solar magnetic field. The models are described and discussed in detail in Ref.~\cite{IceCube:2020bsn}. }
    \label{fig:rd_sun_corr_ssn_inv}
\end{figure}

Absorption of cosmic rays by the Moon and the Sun cause observable deficits (shadows) in the cosmic-ray flux from the corresponding directions. These deficits can be used to verify the directional reconstruction of the detector and, in the case of the Sun, one can study the influence of the solar magnetic field on the observed temporal variation. For such studies, IceCube uses the high-energy (TeV) muons detected in the deep ice. The most recent and most comprehensive study of the Moon and Sun shadows, published in Ref.~\cite{IceCube:2020bsn}, uses seven years of data at median energies of about 50 to 60\,TeV (the estimate is model dependent). While the moon shadow is, as expected, consistent with the geometrical lunar-disk model, a time dependence has been observed for the Sun shadow during the time period from late 2010 until early 2017. This period covers a major part of solar cycle 24 which began in December 2008 and ended in December 2019. The deficit variation from year to year is correlated to the average sunspot number in the respective time period as can clearly be seen in \cref{fig:rd_sun_corr_ssn_inv} (top). The depicted linear fit excludes a constant deficit with a significance of $6.4\,\sigma$. In \cref{fig:rd_sun_corr_ssn_inv} (bottom) the measured relative deficit is compared to the expectations from models of the solar magnetic field. Both models predict a weakening of the shadow in times of high solar activity as it is also visible in the data (see details in Ref.~\cite{IceCube:2020bsn}).

\subsection{IceTop: hybrid detector enhancement} \label{sec:IceTop_enhancement}
Currently, an enhancement program for the surface array is actively pursued to further improve the cosmic-ray science program at the South Pole~\cite{IceCube:2021ydy,IceCube:2023pjc,IceCube:2019yev,IceCube:2021htd}. Detectors of various, complementary types are added to the existing array of IceTop tanks:
\begin{itemize}
\setlength{\itemsep}{3pt}%
    \item scintillator panels,
    \item radio antennas,
    \item and Cherenkov telescopes. 
\end{itemize}	

\begin{figure}[tb]
    \mbox{\hspace{-1em}
    \includegraphics[width=.49\textwidth,trim={0 0 21cm 0},clip]{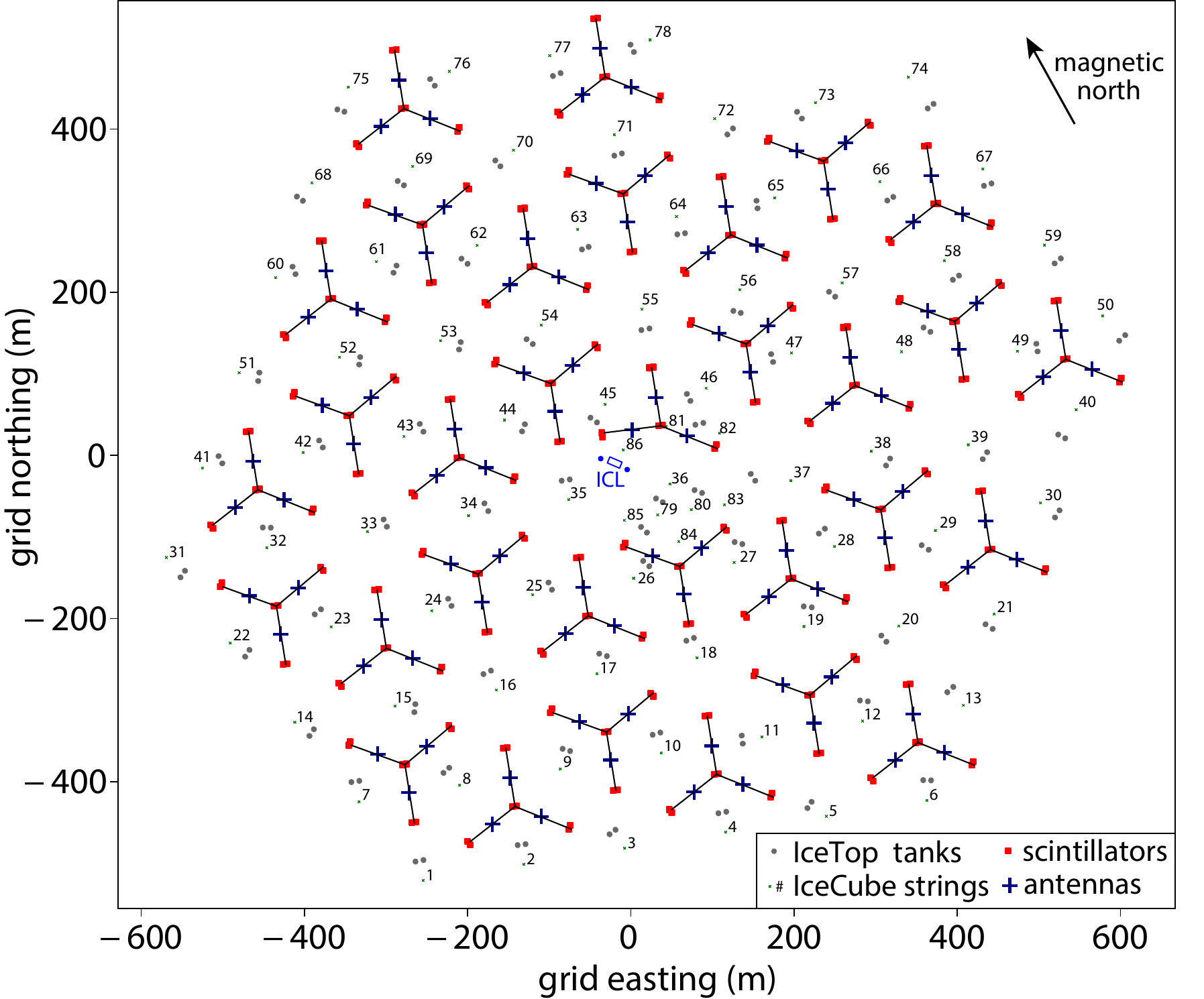}}
    \vspace{-1.5em}
    \caption{Layout of the new scintillator-radio array comprised of 32 detector stations~\cite{IceCube:2023pjc}. Also shown are the locations of the existing IceTop tanks and IceCube strings.}
    \label{fig:IceTopUpgradeMap_StationsLayout_Plan2019}    
\end{figure}

The proposed array layout of scintillator panels and radio antennas is shown in \cref{fig:IceTopUpgradeMap_StationsLayout_Plan2019} with the existing array, IceTop, and the IceCube strings. It is comprised of 32 stations, each station consists of $8$ scintillation panels arranged in pairs, one pair at the center of the station where the local data-acquisition is located, and three pairs at around $70\,\rm{m}$ distance from the center. In addition, three radio antennas with two polarization channels each will be deployed in $35\,\rm{m}$ distance to the center. Currently, a prototype station of scintillator panels and radio antennas~\cite{IceCube:2023pjc}, as well as two Cherenkov telescopes~\cite{IceCube:2023cgq}, are already operating at the South Pole.

With this surface detector enhancement, the following improvements for cosmic-ray physics with IceCube should be achieved within the next decade~\cite{Schroder:2019suq,IceCube:2021okg}:
\begin{itemize}
\setlength{\itemsep}{3pt}%
    \item reduction of systematic uncertainties due to snow coverage of the tanks (antennas and scintillators are  elevated to avoid snow coverage),
    \item  refinement of the cosmic-ray veto for neutrino searches (denser hybrid array),
    \item  adding complementary measurements for composition determinations, like the shower maximum with radio measurements~\cite{IceCube:2023zne},
    \item  opening the path towards a mass-dependent measurement of the cosmic-ray anisotropy,
    \item  making searches for PeV gamma-rays more competitive (until now only upper limits could be given~\cite{IceCube:2019scr}),
    \item  adding more and complementary input for tuning hadronic interaction models.
\end{itemize}

Furthermore, the surface enhancement also serves as a prototype for the surface array planned for the future extension IceCube-Gen2 \cite{IceCube-Gen2:2020qha,IceCube-Gen2:2021aek,ColemanICRC}, which will have an approximately $8\,\rm{km}^2$ surface coverage. This next-generation observatory will collect high-quality data with unprecedented statistics and thereby continue the longstanding Antarctic science program to further improve our understanding of high-energy cosmic rays.

%% file: sec5.tex
Over more than five decades, a variety of experiments at the South Pole have provided important information on the nature and origin of cosmic rays. Since the 1950s, neutron monitors provide ground-based measurements of solar cosmic rays in the few GeV energy regime by detecting the hadronic component of the atmospheric cascade, allowing detailed studies of the Sun and its influence on the heliosphere, as well as the geomagnetic field. With strong contributions from the Bartol Research Institute the first detector in Antarctica was established in 1960 at McMurdo Station which was relocated to the South Pole in 1964.

The idea of constructing a high-energy air-shower array at the South Pole was first considered in the early 1980s by Michael Hillas and others. The high elevation allows for a low energy threshold for large air-shower arrays. In mid-1986, a Bartol-Leeds collaboration was formed with major contributions from Tom Gaisser and his group to operate the SPASE air-shower array with deployment beginning in late 1987. In addition, a string with four optical sensors was deployed at a depth of $800\,\rm{m}$ near the center of the SPASE-1 array. This hybrid setup successfully allowed for the observation of coincidences of muons in the deep ice with air showers detected at the surface. A comparison of the observed muon coincidence rates with the expected rates obtained from simulations provided an indirect measurement of the ice transparency, demonstrating the feasibility of cosmic neutrino detection in the Antarctic ice.

These observations motivated the construction of AMANDA-A at depths between $800$ and $1000\,\rm{m}$ which obtained first results in 1994. It was realized that the transparency of the ice improves with depth. This lead to the construction of AMANDA-B at deeper depths, between $1520$ and $1900\,\rm{m}$, and SPASE-II. The main result from the analysis of coincident data was a measurement of the mean mass composition of cosmic rays with energies between $500\,\rm{TeV}$ and $5\,\rm{PeV}$. These observations laid the foundations for the construction of the IceCube Neutrino Observatory which started in 2004.

IceCube has demonstrated the strong scientific capabilities of a hybrid, large-scale particle detector at the South Pole. The main focus of the detector is the discovery and measurement of astrophysical neutrinos. However, the successful coincident operation of the AMANDA and SPASE detectors has shown the large potential for cosmic-ray physics. 

The Bartol group under the leadership of Tom Gaisser played a crucial role in the development and realization of the cosmic-ray program of the IceCube collaboration. In particular, the opportunity for coincident measurements of air showers at the surface and muons in the deep ice was of large interest. It enables analyses of the mass composition of cosmic rays from a few PeV up to around $1\,\rm{EeV}$, for example, providing important information on the nature of Galactic cosmic rays. In addition, measurements of GeV-muons at the surface and TeV-muons in the deep ice provide unique information on particle production in air showers. 

Studies of the arrival direction of high-energy muons enable measurements of the cosmic-ray anisotropy in the Southern Hemisphere with large statistics over a wide energy range. These analyses provide the dipole amplitude and phase, as well as a measurement of the small scale structures of the observed anisotropy. Studies of the correlations of the observed structures with features of the interstellar magnetic field yield important information about the potential sources and propagation of cosmic rays. 

Motivated by the successful operation of the IceCube Neutrino Observatory for more than a decade, with a large variety of important science results, an upgrade of the existing detector is currently in progress. This includes an enhancement for the surface array which will further improve cosmic-ray measurements at the South Pole. The upgrade of the IceCube Observatory will continue the longstanding Antarctic science program and promises an exciting scientific future. Thereby, the existing and future measurements at the South Pole will continue the ground-breaking research conveyed by Tom K. Gaisser and his colleagues.